\documentclass[prx,twocolumn,showpacs,floatfix,superscriptaddress,amsmath,amsfonts,amssymb,preprintnumbers,citeautoscript,aps,longbibliography]{revtex4-1}
\pdfoutput=1
\usepackage[T1]{fontenc}\usepackage[latin1]{inputenc}
\usepackage{dcolumn,graphicx,color,booktabs,microtype,afterpage} \graphicspath{{fig/}}
\usepackage[charter,greekuppercase=italicized]{mathdesign}\usepackage{sidecap}
\usepackage[colorlinks,plainpages=false,linkcolor=blue,urlcolor=blue,citecolor=blue,pdfpagemode=UseNone,pdfstartview=FitBH]{hyperref}

\begin{document}
\title{Lattice dynamics in the double-helix antiferromagnet FeP}
\author{A. S. Sukhanov}
\affiliation{Max Planck Institute for Chemical Physics of Solids, D-01187 Dresden, Germany}
\affiliation{Institut f{\"u}r Festk{\"o}rper- und Materialphysik, Technische Universit{\"a}t Dresden, D-01069 Dresden, Germany}
\author{S. E. Nikitin}
\affiliation{Max Planck Institute for Chemical Physics of Solids, D-01187 Dresden, Germany}
\affiliation{Paul Scherrer Institute (PSI), CH-5232 Villigen, Switzerland}
\author{M. S. Pavlovskii}
\affiliation{Kirensky Institute of Physics, Siberian Branch, Russian Academy of Sciences, Krasnoyarsk 660036, Russian Federation}
\author{T. C. Sterling}
\affiliation{Department of Physics, University of Colorado at Boulder, Boulder, CO 80309, United States}
\author{N.~D.~Andryushin}
\affiliation{Kirensky Institute of Physics, Siberian Branch, Russian Academy of Sciences, Krasnoyarsk 660036, Russian Federation}
\author{A.~S.~Cameron}
\affiliation{Institut f{\"u}r Festk{\"o}rper- und Materialphysik, Technische Universit{\"a}t Dresden, D-01069 Dresden, Germany}
\author{Y.~V.~Tymoshenko}
\affiliation{Institut f{\"u}r Festk{\"o}rper- und Materialphysik, Technische Universit{\"a}t Dresden, D-01069 Dresden, Germany}
\author{H.~C.~Walker}
\affiliation{ISIS Facility, STFC, Rutherford Appleton Laboratory,
Didcot, Oxfordshire OX11-0QX, United Kingdom}
\author{I.~V.~Morozov}
\affiliation{Leibniz Institute for Solid State and Materials Research Dresden, D-01069 Dresden, Germany}
\affiliation{Lomonosov Moscow State University, 119991 Moscow, Russia}
\author{I.~O.~Chernyavskii}
\affiliation{Leibniz Institute for Solid State and Materials Research Dresden, D-01069 Dresden, Germany}
\affiliation{Lomonosov Moscow State University, 119991 Moscow, Russia}
\author{S.~Aswartham}
\affiliation{Leibniz Institute for Solid State and Materials Research Dresden, D-01069 Dresden, Germany}
\author{D.~Reznik}
\affiliation{Department of Physics, University of Colorado at Boulder, Boulder, CO 80309, United States}
\author{D.~S.~Inosov}
\affiliation{Institut f{\"u}r Festk{\"o}rper- und Materialphysik, Technische Universit{\"a}t Dresden, D-01069 Dresden, Germany}
\affiliation{W\"{u}rzburg-Dresden Cluster of Excellence on Complexity and Topology in Quantum Matter\,---\,\textit{ct.qmat}, TU Dresden, D-01069 Dresden, Germany}
\begin{abstract}

We present a comprehensive investigation of lattice dynamics in the double-helix antiferromagnet FeP by means of high-resolution time-of-flight neutron spectroscopy and ab-initio calculations. Phonons can hybridize with the magnetic excitations in noncollinear magnets to significantly influence their properties. We observed a rich spectrum of phonon excitations, which extends up to $\sim$50~meV. We performed detailed analysis of the observed and calculated spectra for all high-symmetry points and high-symmetry directions of the Brillouin zone. We show that the DFT calculations quantitatively capture the essential features of the observed phonons, including both dispersions and scattering intensities. By making use of the detailed intensity comparison between the theory and the data, we were able to identify displacement vectors for the majority of the observed modes. The overall excellent agreement between the DFT predictions and the experimental results breaks down for the lowest mode at the $Y$-point, whose energy is lower than calculated by $\sim$13\%. The present study provides vital information on the lattice dynamics in FeP and demonstrates applicability of the DFT to novel pressure-induced phenomena in related materials, such as MnP and CrAs.

\end{abstract}

\maketitle

\section{Introduction}

Transition-metal monophosphates $A$P ($A$ = Fe, Mn) are metallic binary compounds that attracted significant attention due to their magnetic and electronic properties. Both materials are itinerant helimagnets featuring an unusual ``double helix'' type of magnetic ordering with the propagation vector aligned along the $c$-axis of the orthorhombic crystal structure~\cite{felcher1971magnetic, haggstrom1982mossbauer,felcher1966magnetic}. In addition, an unusual orbital density-wave ordering was recently observed in MnP by means of resonant inelastic x-ray scattering. The orbital ordering has half the period of the magnetic helix and shows a similar temperature dependence which implies a coupling between the two orders~\cite{pan2019intertwined}. This feature is reminiscent of the nematic orbital order in the iron-based superconductors LaFeAsO and BaFe$_{2-x}$Co$_x$As$_2$~\cite{PhysRevB.82.104519,Chu824,Yi6878}.

Application of hydrostatic pressure in MnP suppresses the magnetic transition temperature down to 0~K and induces a quantum phase transition toward a nonmagnetic state. Pressure-induced variation of the magnetic spiral structure of MnP has been studied by nonresonant magnetic x-ray diffraction~\cite{Wang2016}, muon-spin rotation~\cite{PhysRevB.93.180509}, and neutron diffraction~\cite{PhysRevB.93.100405}. It was shown that the ambient-pressure incommensurate helical structure shortens, becomes commensurate, and rotates its propagation vector to the $b$-axis at a pressure of $\sim$2~GPa. This suggests a delicate balance of competing magnetic interactions~\cite{Wang2016,PhysRevB.93.180509,PhysRevB.93.100405}. Surprisingly, an unconventional superconducting phase was observed below $T_{\mathrm{c}} \approx 1$~K at $P \approx 8$~GPa, very close to the quantum critical point, making MnP the first ever observed Mn-based superconductor~\cite{cheng2015pressure}.

Even though the electronic~\cite{xu2017first, de1995local,PhysRevMaterials.3.095004,Han2018,Nozue2001192}, transport~\cite{westerstrandh1977magnetic, kawakatsu2019haas, yin2019electrical,PhysRevB.86.180404}, and magnetic~\cite{felcher1971magnetic, haggstrom1982mossbauer, felcher1966magnetic, yano2018suppression, zhigadlo2017growth} properties of MnP and FeP have been intensively studied by different experimental techniques, there are no reports on the structural dynamics in these materials. Structural vibrations---phonons---are known to serve as a gluing mechanism for Cooper pairing in BCS theory of superconductivity. Moreover,  phonons can hybridize with magnetic excitations in noncollinear magnets~\cite{toth2016electromagnon, oh2016spontaneous}, which may significantly influence their properties. In the isostructural material MnAs (in its orthorhombic phase, which is stable between 315 and 393~K), giant coupling between a phonon soft mode and magnetic moments plays a crucial role in the magnetostructural phase transition, as evidenced by first-principles calculations~\cite{lazewski2010phonon,PhysRevB.83.054108,PhysRevB.74.024429}. Thus, knowledge of phonon dynamics can provide essential information on unusual physics in these materials.

Unlike MnAs, which is paramagnetic in its orthorhombic phase, another 3d-metal pnictide isostructural to FeP and MnP---CrAs---orders in a double-helix magnetic structure.~\cite{BOLLER19711699}. The magnetic and lattice degrees of freedom were also found to be coupled in this material. The magnetic transition in CrAs is accompanied by a first-order isostructural transition which manifests itself by a large abrupt expansion of the $b$-axis and a slight reduction of $a$ and $c$-axes~\cite{doi:10.1143/JPSJ.30.1319}. Similarly to MnP, pressure-induced superconductivity was discovered in CrAs in the vicinity of the helical phase~\cite{Wu2014,PhysRevB.98.024520,PhysRevLett.114.117002,PhysRevX.8.031017,Park2019}, and the magnetic propagation vector in CrAs undergoes the reorientation from $c$ to $b$ upon  application of moderate pressure~\cite{PhysRevB.93.060503}.

In FeP, the magnetic, lattice, and electronic degrees of freedom seem to be coupled to a lesser extent. Thus FeP can be considered as a model system for the phonon dynamics in the 3d-metal monopnictides family. In this work, we present the results of a comprehensive investigation of structural dynamics in FeP. We combine thermal-neutron time-of-flight spectroscopy and $\textit{ab-initio}$ phonon calculations in order to fully characterize the phonons over the entire Brillouin zone. The remainder of the manuscript is organized as follows: Section~\ref{Sec:II} describes the experimental details and the first-principles calculations. In Section~\ref{Sec:III} we discuss results of the measurements and compare the obtained neutron spectra with the calculations. The analysis is first presented for all high-symmetry points in the Brillouin zone, where the majority of lattice vibrations are resolved and identified. Then, the data on the phonon dispersions along high-symmetry directions are shown. In section~\ref{Sec:IV}, we summarize the results.

\section{Methods}\label{Sec:II}

\begin{figure}[t]
        \begin{minipage}{0.99\linewidth}
        \center{\includegraphics[width=1\linewidth]{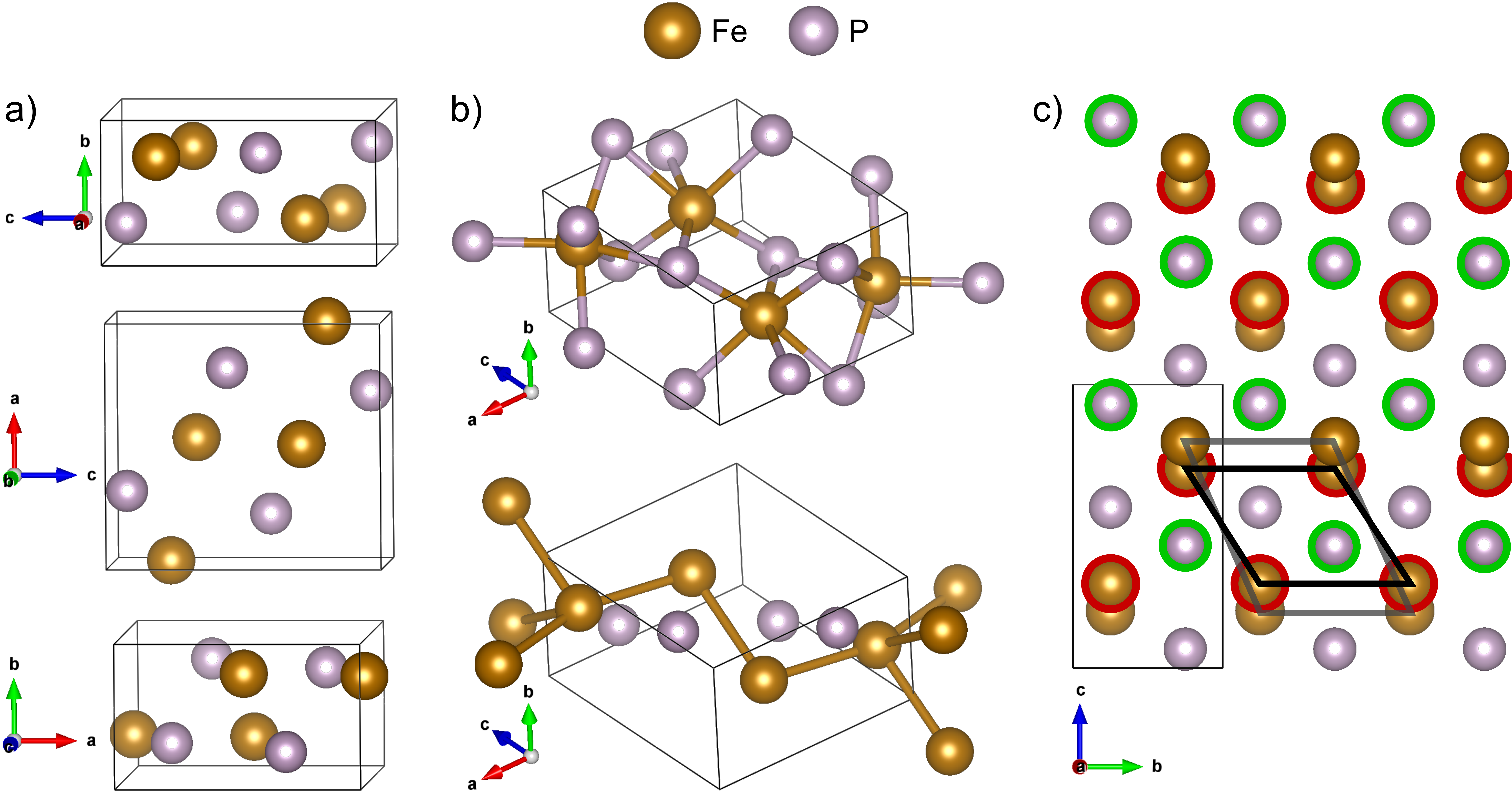}}
        \end{minipage}
        \caption{(color online). The crystal structure of FeP. (a) The unit cell shown along different crystal axes. (b) The shortest Fe-P and Fe-Fe bonds are shown within the unit cell. (c) The projection of the crystal structure onto the $bc$-pane, which highlights the relation to the distorted hexagonal structure. Red and green circles label the Fe and P atoms that form the buckled triangular layers distorted along $c$. Black tetragons highlight the distortion of the Fe layers.}
        \label{ris:fig0}
\end{figure}

The crystal structure of FeP is depicted in Figs.~\ref{ris:fig0}(a)--\ref{ris:fig0}(c). It is described by the space group $Pnma$ (no. 62) with the lattice parameters $a = 5.197$~\AA, $b = 3.099$~\AA, and $c = 5.794$~\AA~ at room temperature~\cite{felcher1971magnetic,Chernyavskii}. Both Fe and P occupy the $4c$ Wyckoff position with the parameters $x = 0.002$ and $z = 0.200$ for Fe and $x = 0.191$ and $z = 0.569$ for P. It is worth noting that the $Pnma$ crystal structure of FeP can be considered as a distorted hexagonal NiAs-type structure ($P6_3/mmc$ space group), where the distorted hexagonal layers of Fe ions are stacked along the orthorhombic $a$-axis. Figure~\ref{ris:fig0}(c) depicts the crystal structure of FeP as viewed from [100]. Red circles highlight the Fe ions that have approximately the same coordinate along the $a$-axis, namely, $x_1 = 0.048$ and $x_2 = 0.502$. The highlighted sites form the buckled triangular layers that are contracted along the $c$-axis such that the isosceles triangle of neighbouring ions has the base angle of $\sim$56$^{\circ}$. The next Fe layer is then elongated along $c$ with a $\sim$66$^{\circ}$ angle. Green circles in Fig.~\ref{ris:fig0}(c) highlight the P ions that form a triangular layer with a larger buckling along $a$ (the relative difference of the $x$ coordinates for P layers $x_1 - x_2 = 0.118$), but with a smaller distortion along $c$.

\subsection{Experimental details}

High quality single crystals of FeP up to 500 mg in mass and 80 mm$^3$ in volume were grown by chemical vapor transport with iodine as a transport agent. The quality of the crystals were confirmed by means of EDX, XRD, magnetic, transport, high resolution TEM and single-crystal neutron diffraction experiments~\cite{Chernyavskii}. The temperature dependence of the resistivity shows the typical metallic behavior with a residual resistivity ratio RRR = 566. Below the the N\'{e}el temperature $T_{\text{N}} = 119.3$~K, analysis of the neutron diffraction data yields an incommensurate magnetic structure with the propagation vector $\mathbf{Q} =(0, 0, \pm \delta)$, where $\delta \approx 0.2$~\cite{felcher1971magnetic,Chernyavskii}. For neutron spectroscopy measurements, we used a single crystal with a mass of $\sim$0.5~g.

The experiment was conducted at the thermal-neutron direct-geometry time-of-flight spectrometer Merlin~\cite{BEWLEY20061029} located at the ISIS Neutron and Muon Source (Didcot, UK). The sample was oriented with its [100] axis vertical, thus the ($0KL$) reciprocal plane was horizontal throughout the measurements. In addition to a large 2$\theta$ angular range of 135$^{\circ}$ in the horizontal plane, the detector banks of Merlin also allow for detection of scattered neutrons up to 30$^{\circ}$ in 2$\theta$ in the vertical plane. This enabled us to collect the data on the elemental excitations for all principal directions in reciprocal space within a single experimental setup. The incident neutron energy $E_{\text{i}}$ was set to 60~meV to achieve a reasonable experimental resolution while sampling a sufficiently large volume the momentum-energy space. The trade-off between the resolution and intensity was optimized by setting the chopper frequency to 450~Hz. In this setup, the resulting energy resolution at the elastic line is estimated as $\sim$3~meV. An additional dataset with a reduced counting time was collected at $E_{\text i} = 40$~meV. During the measurements, the crystal was gradually rotated over 70$^{\circ}$ around the [100] axis with a step of 0.5$^{\circ}$. To minimize anharmonic effects on the phonon spectra, all the measurements were performed at $T = 6$~K.

The collected data were reduced and analysed using the \textsc{horace} software~\cite{EWINGS2016132}. A symmetrization procedure was applied during data reduction, which means that the data from equivalent $\textbf{Q}$-directions in momentum space [for example, $(H00)$ and $(-H00)$] were averaged to improve statistics. Thus, the covered $\textbf{Q}$-space was folded down to a 90$^{\circ}$ sector in the ($0KL$) plane, which is irreducible for the orthorombic system. To plot the $I(Q,E)$ intensity colormaps for different momentum directions, the data were integrated over $\pm 0.2$~\AA$^{-1}$ along the perpendicular momenta. The scattering intensity profiles in the high-symmetry points of the BZ ($I(E)$) were integrated over $\pm 0.1$~r.l.u. along each principal momentum direction [$(H00)$, $(0K0)$, and $(00L)$].

Phonon energies were obtained with the Phonon Explorer software, which works with unsymmetrized raw data~\cite{https://github.com/dmitryr1234/phonon-explorer}. It effectively combines statistics from all Brillouin zones where the phonon of interest has appreciable intensity as described in Sec. IIID.

\subsection{First-principles calculations}

Lattice dynamics calculations were carried out by means of the projector-augmented wave (PAW) method~\cite{PhysRevB.59.1758} and density functional theory (DFT) as implemented in the \textsc{vasp} software~\cite{PhysRevB.54.11169,KRESSE199615}. The generalized gradient approximation (GGA) functional with Perdew-Burke-Ernzerhof (PBE) parametrization~\cite{PhysRevLett.77.3865} was used. The plane-wave cutoff was set to 600 eV. An $8\times13\times3$ (Monkhorst-Pack scheme)~\cite{PhysRevB.13.5188} $k$-point mesh was used for Brillouin zone integration. The following electronic configurations was chosen: $3d^{7}4s^{1}$ for Fe, and $3p^{3}3s^{2}$ for P. Magnetic moments of Fe ions were taken into account by performing spin polarized calculations. The phonons were calculated by constructing a supercell ($2\times2\times2$) and calculating the force constants as implemented in \textsc{phonopy}~\cite{TOGO20151}. Classification of vibrational modes was performed by means of \textsc{phonopy}. The notation used for the irreducible representations corresponds to \textsc{isotropy}~\cite{IsoTrop}.

In an INS experiment, the dynamic structure factor $S(Q,E)$ is measured, which takes the following form for the lattice vibrations:

\begin{equation}
\begin{split}
S(\mathbf{Q},E) \propto \sum_{s,\mathbf{q},\tau} \frac{1}{E_s\left(\mathbf{q}\right)} \left\vert  \sum_d \frac{\overline{b_d}}{\sqrt{M_d}} e^{i\mathbf{Q} \cdot \mathbf{r}_d} e^{-W_d}  \left( \mathbf{Q} \cdot \epsilon_{d,s} \right) \right\vert^2 \\ \times \left\langle n_s+1 \right\rangle \delta\left( E_s\left(\mathbf{q}\right)-E\right) \delta\left(\mathbf{Q}-\mathbf{q}-\tau \right),
\end{split}
\label{eq:eq1}
\end{equation}
where $E$ and $\epsilon_{d,s}$ are the energy and the polarization of the phonon with the mode index $s$ and the wave-vector $\mathbf{q}$, $\tau$ is a reciprocal lattice vector, $\mathbf{r}_d$ is the position vector for the atom $d$ in the unit cell, $M_d$ its mass, and $\overline{b_d}$ its coherent cross-section; $W_d$ is the Debye-Waller factor, and $n_s$ is the Bose-Einstein distribution. The INS intensity was simulated using the \textsc{oclimax} software~\cite{doi:10.1021/acs.jctc.8b01250} which uses vibrational frequencies and polarization vectors from first-principles calculations (\textsc{phonopy} in this work) as the input. The simulated spectra were convolved with a Gaussian function with the standard deviation $\sigma = 2$~meV to model the experimental resolution.

\section{Results}\label{Sec:III}

\subsection{Dynamic structure factor}

\begin{figure}[t]
        \begin{minipage}{0.99\linewidth}
        \center{\includegraphics[width=1\linewidth]{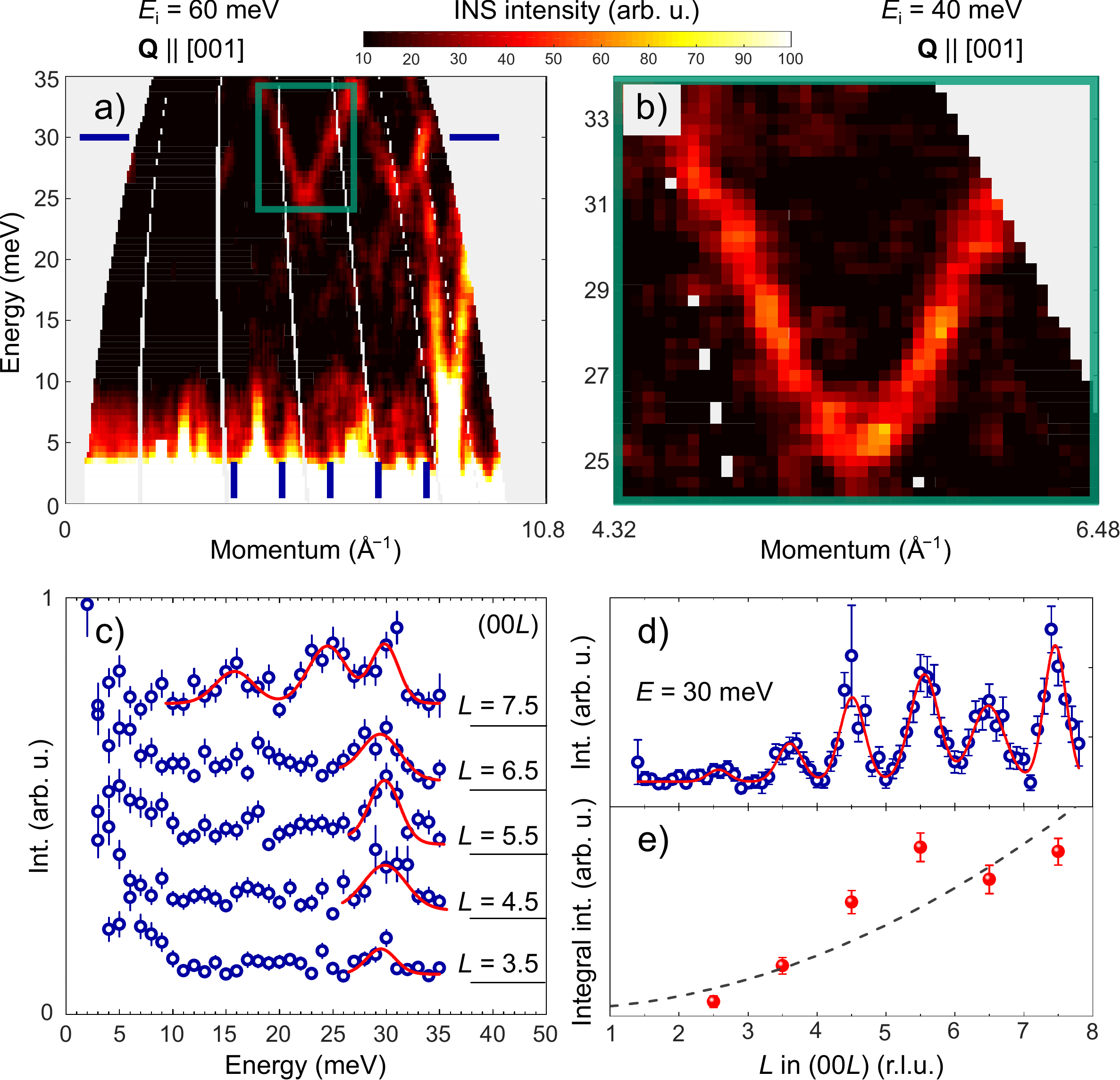}}
        \end{minipage}
        \caption{(color online). (a)~INS intensity as a function of the neutron energy loss and momentum transfer along the $c$-axis of the crystal. (b)~A closer view of the observed dispersion in a narrower $Q$-$E$ range is highlighted by the frame in (a). (c)~The scattering intensity as a function of the energy transfer at different momenta along $(00L)$ (in r.l.u.). The corresponding momenta are marked as the vertical bars in (a). (d)~A constant-energy momentum cut [$E$-position is marked by the horizontal bar in (a)] at 30~meV along the $(00L)$ direction. The symbols are experimental data, the solid line is a fit by a linear combination of Gaussian functions. (e)~The integrated intensity (the area of a Gaussian function) of each fitted peak in (d) as a function of the peak position. The dotted curve is a fit to a quadratic function.}
        \label{ris:fig1}
\end{figure}

\begin{figure}[t]
        \begin{minipage}{0.99\linewidth}
        \center{\includegraphics[width=1\linewidth]{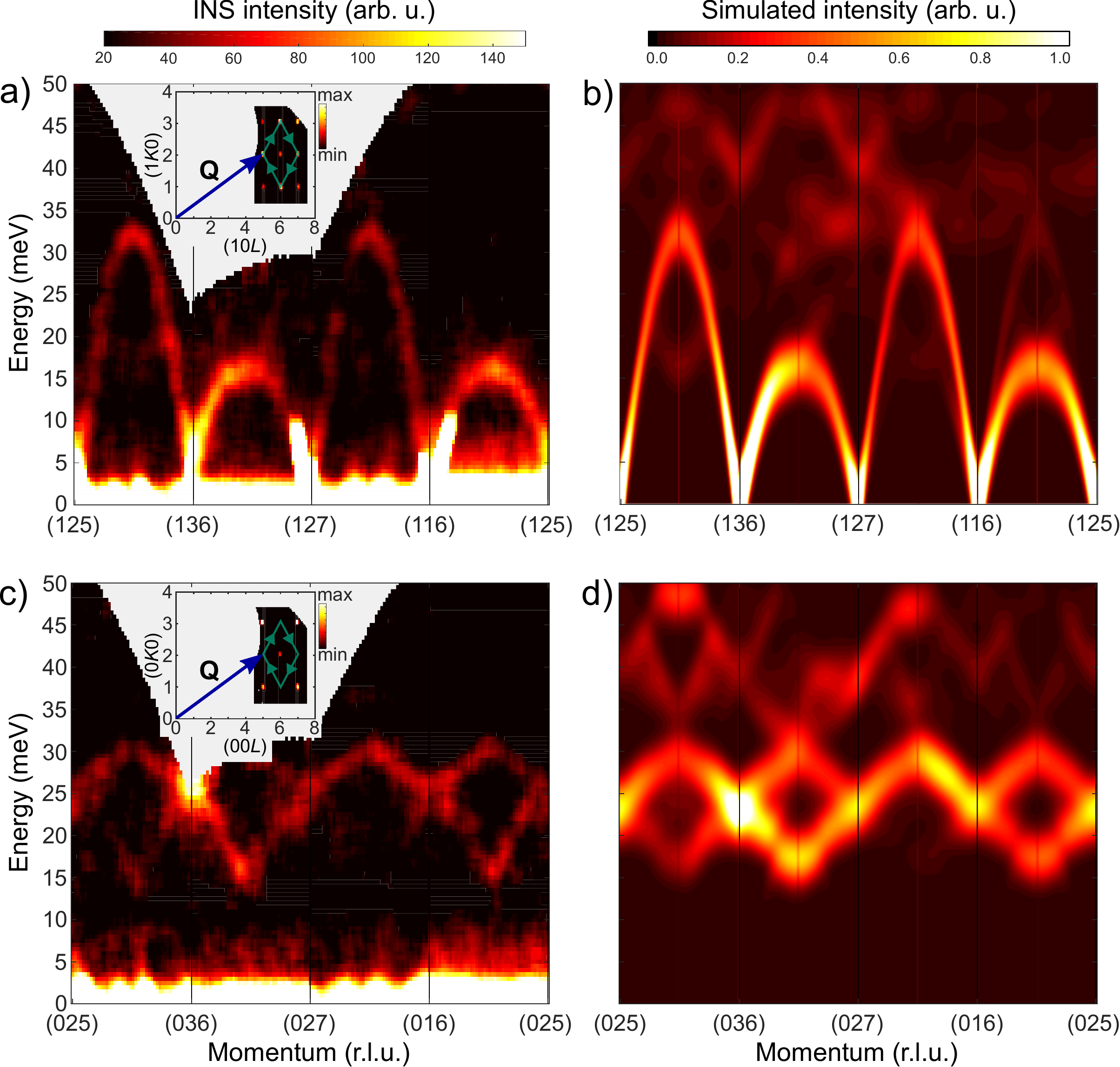}}
        \end{minipage}
        \caption{(color online). Momentum-energy cuts through the INS data demonstrating the variation of the phonon intensity due to the phonon polarization and the structure factor in the ($1KL$) and ($0KL$) reciprocal planes~[(a) and (c)]. (b) and (d)~The simulated phonon dynamical structure factor. The traversed path in reciprocal space is depicted in the insets to (a) and (c).}
        \label{ris:fig2}
\end{figure}

\begin{figure}[t]
        \begin{minipage}{0.99\linewidth}
        \center{\includegraphics[width=1\linewidth]{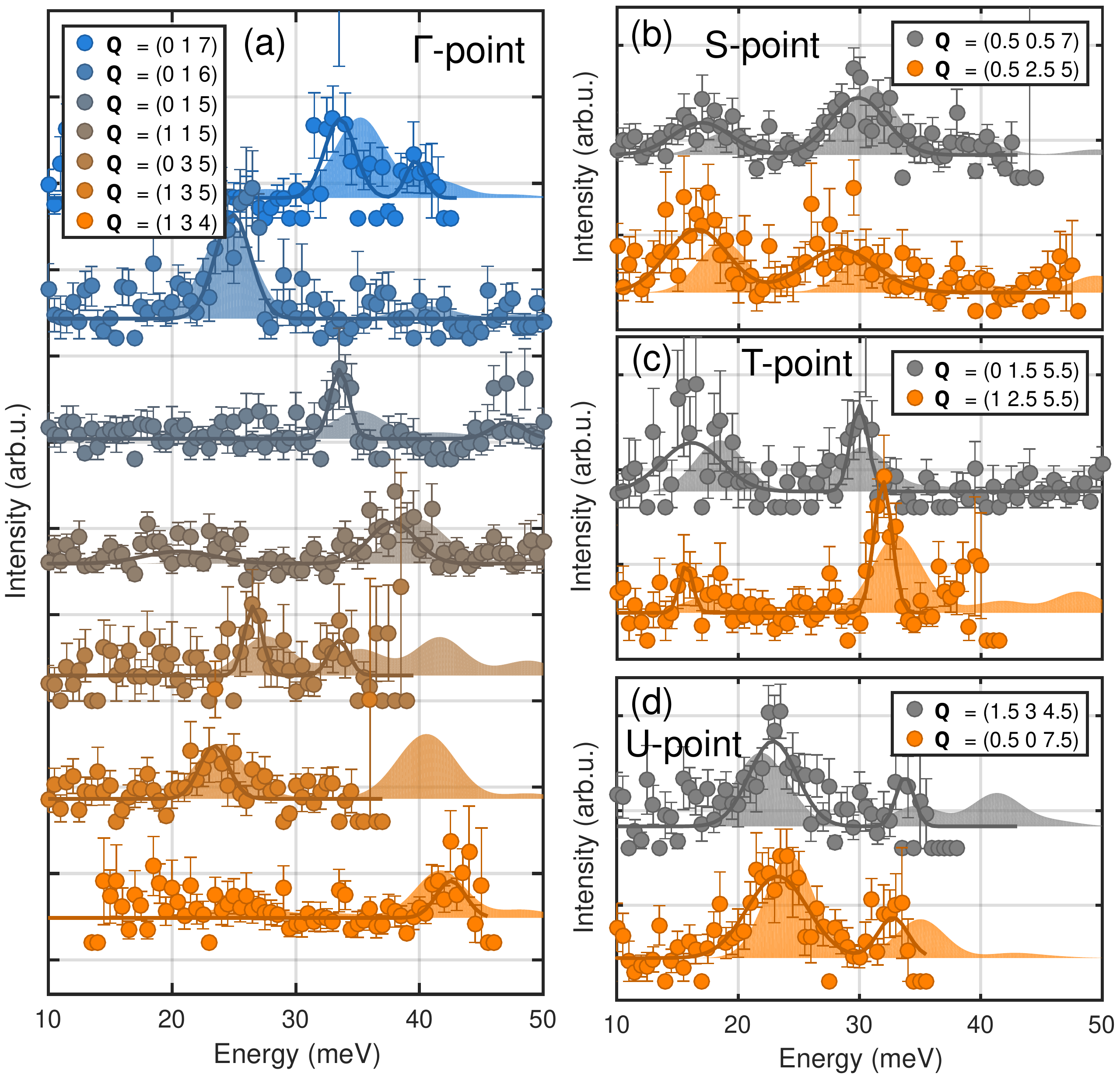}}
        \end{minipage}
        \caption{(color online). Comparison between the experimental data (points) and results of the \textit{ab initio} calculations (shaded area) at high symmetry points of the BZ adjacent to different reciprocal-lattice points, (a) the $\Gamma$-point, (b) $S$, (c) $T$, (d) $U$. Solid lines are Gaussian fits to the experimental data.}
        \label{ris:fig3}
\end{figure}

\begin{figure}[t]
        \begin{minipage}{0.99\linewidth}
        \center{\includegraphics[width=1\linewidth]{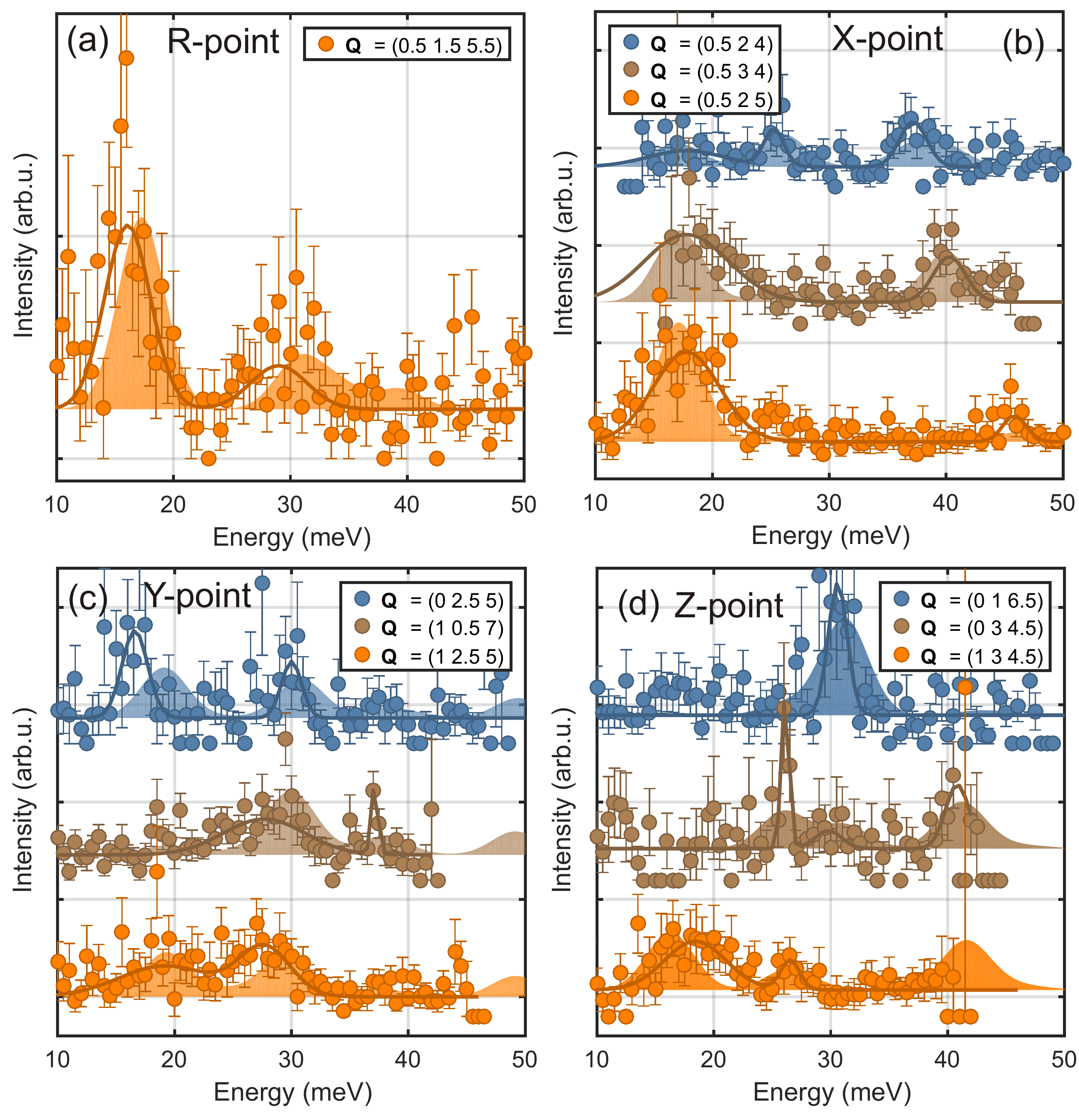}}
        \end{minipage}
        \caption{(color online). Comparison between the experimental data (points) and results of the \textit{ab initio} calculations (shaded area) at high symmetry points of the BZ adjacent to different reciprocal-lattice points, (a) the $R$-point, (b) $X$, (c) $Y$, (d) $Z$. Solid lines are Gaussian fits to the experimental data.}
        \label{ris:fig4}
\end{figure}

\begin{table}[!t]
\caption{Comparison between the experimentally determined phonon energies and the results of the \textit{ab initio} calculations at all the high-symmetry points of the BZ. The reciprocal points at which a particular phonon mode was observed are listed in the first column.}
\label{tab:tab1}
\begin{tabular}{@{}l@{}l|c|c|c@{}}
\toprule
\multicolumn{2}{c|}{$\mathbf{Q}$-point} & $E_{\mathrm{exp}}$~(meV)  & $E_{\mathrm{calc}}$~(meV) & Symmetry  \\
\midrule
$\Gamma$ & $(1~3~5)$                          &  23.4(4)                     & 23.86                         & $B_{1g}^1$ \\
& $(0~1~6)$                                 &  24.9(2)                     & 24.74                         & $B_{3u}^1$ \\
& $(0~3~5)$                                 &  26.5(2)                     & 27.60                         & $B_{3g}^1$ \\
& $(0~1~5)$                                 &  33.5(2)                     & 35.27                         & $A_g^2$ \\
& $(1~1~5)$                                 &  37.8(4)                     & 39.46                         & $B_{2g}^1$ \\
& $(0~1~7)$                                 &  39.6(4)                     & 41.49                         & $A_g^3$ \\
& $(1~3~4)$                                 &  42.5(4)                     & 41.56, 41.88                         & $B_{1g}^2$, $B_{2u}^1$ \\
& $(0~1~5)$                                 &  47.1(7)                     & 49.02                         & $A_g^4$ \\
\midrule
$S$ & $\left(\frac{1}{2}~\frac{1}{2}~7\right)$   &  16.9(5)                    & 18.92                         & $S_1$+$S_2$ \\
    & $\left(\frac{1}{2}~\frac{1}{2}~7\right)$     &  29.9(3)                     & 30.83                         & $S_1$+$S_2$\rule{0pt}{1.2em} \\
\midrule
$T$ & $(0~\frac{3}{2}~\frac{11}{2})$            &  16.1(6)                     & 17.40                         & $T_2$ \\
& $(0~\frac{3}{2}~\frac{11}{2})$            &  29.9(2)                     & 30.27                         & $T_1$\rule{0pt}{1.2em}  \\
& $(1~\frac{5}{2}~\frac{11}{2})$            &  31.9(1)                     & 33.10                         & $T_2$\rule{0pt}{1.2em}  \\
\midrule
$U$ & $(\frac{3}{2}~3~\frac{9}{2})$           &  22.9(3)                     & 21.53                         &  $U_2^+$ + $U_3^+$ \\
& $(\frac{1}{2}~0~\frac{15}{2})$            &  23.2(3)                     & 24.35                         & $U_2^-$ + $U_3^-$\rule{0pt}{1.2em}  \\
& $(\frac{1}{2}~0~\frac{15}{2})$            &  32.7(7)                     & 35.08                         & $U_2^-$ + $U_3^-$\rule{0pt}{1.2em}  \\
& $(\frac{3}{2}~3~\frac{9}{2})$             &  33.7(7)                     & 34.94                         & $U_1^+$ + $U_4^+$\rule{0pt}{1.2em}  \\
\midrule
$X$ & $(\frac{1}{2}~2~5)$                       &  17.8(3)                     & 16.74, 17.08                         & $X_1$, $X_2$ \\
& $(\frac{1}{2}~2~4)$                       &  25.2(3)                     & 25.86                         & $X_2$\rule{0pt}{1.2em}  \\
& $(\frac{1}{2}~2~4)$                       &  37.1(4)                     & 37.16, 39.84                  & $X_1$, $X_1$\rule{0pt}{1.2em}  \\
& $(\frac{1}{2}~3~4)$                       &  40.2(3)                     & 40.16                         & $X_2$\rule{0pt}{1.2em}  \\
& $(\frac{1}{2}~2~5)$                       &  45.8(3)                     & 47.28                         & $X_1$\rule{0pt}{1.2em}  \\
\midrule
$Y$ & $(0~\frac{5}{2}~5)$                     &  16.6(2)                     & 19.20                         & $Y_2$ \\
& $(1~\frac{5}{2}~5)$                       &  19(1)                       & 19.41                         & $Y_1$\rule{0pt}{1.2em}  \\
& $(1~\frac{5}{2}~5)$                       &  27.6(6)                     & 29.80                         & $Y_2$\rule{0pt}{1.2em}  \\
& $(0~\frac{5}{2}~5)$                       &  30.0(3)                     & 31.05                         & $Y_1$\rule{0pt}{1.2em}  \\
& $(1~\frac{1}{2}~7)$                       &  37.1(1)                     & 37.98                         & $Y_2$\rule{0pt}{1.2em}  \\
\midrule
$Z$ & $(1~3~\frac{9}{2})$                     &  18.4(4)                     & 16.53                         & $Z_2$ \\
& $(0~3~\frac{9}{2})$                       &  26.11(8)                    & 26.06                         & $Z_2$\rule{0pt}{1.2em}  \\
& $(1~3~\frac{9}{2})$                       &  26.5(3)                     & 26.06, 26.89                         &$Z_2$, $Z_1$\rule{0pt}{1.2em}  \\
& $(0~1~\frac{13}{2})$                      &  30.6(2)                     & 31.36                         & $Z_1$\rule{0pt}{1.2em}  \\
& $(0~3~\frac{9}{2})$                       &  40.8(4)                     & 40.99, 41.50                         & $Z_2$, $Z_2$\rule{0pt}{1.2em}  \\
\midrule
$R$ & $(\frac{1}{2}~\frac{3}{2}~\frac{11}{2})$&  16.0(3)                     & 17.23                         & $R_1$+$R_2$ \\
& $(\frac{1}{2}~\frac{3}{2}~\frac{11}{2})$  &  29(1)                       & 30.72                         & $R_1$+$R_2$\rule{0pt}{1.2em}  \\
\bottomrule
\end{tabular}
\end{table}

\begin{figure}[t]
        \begin{minipage}{0.99\linewidth}
        \center{\includegraphics[width=1\linewidth]{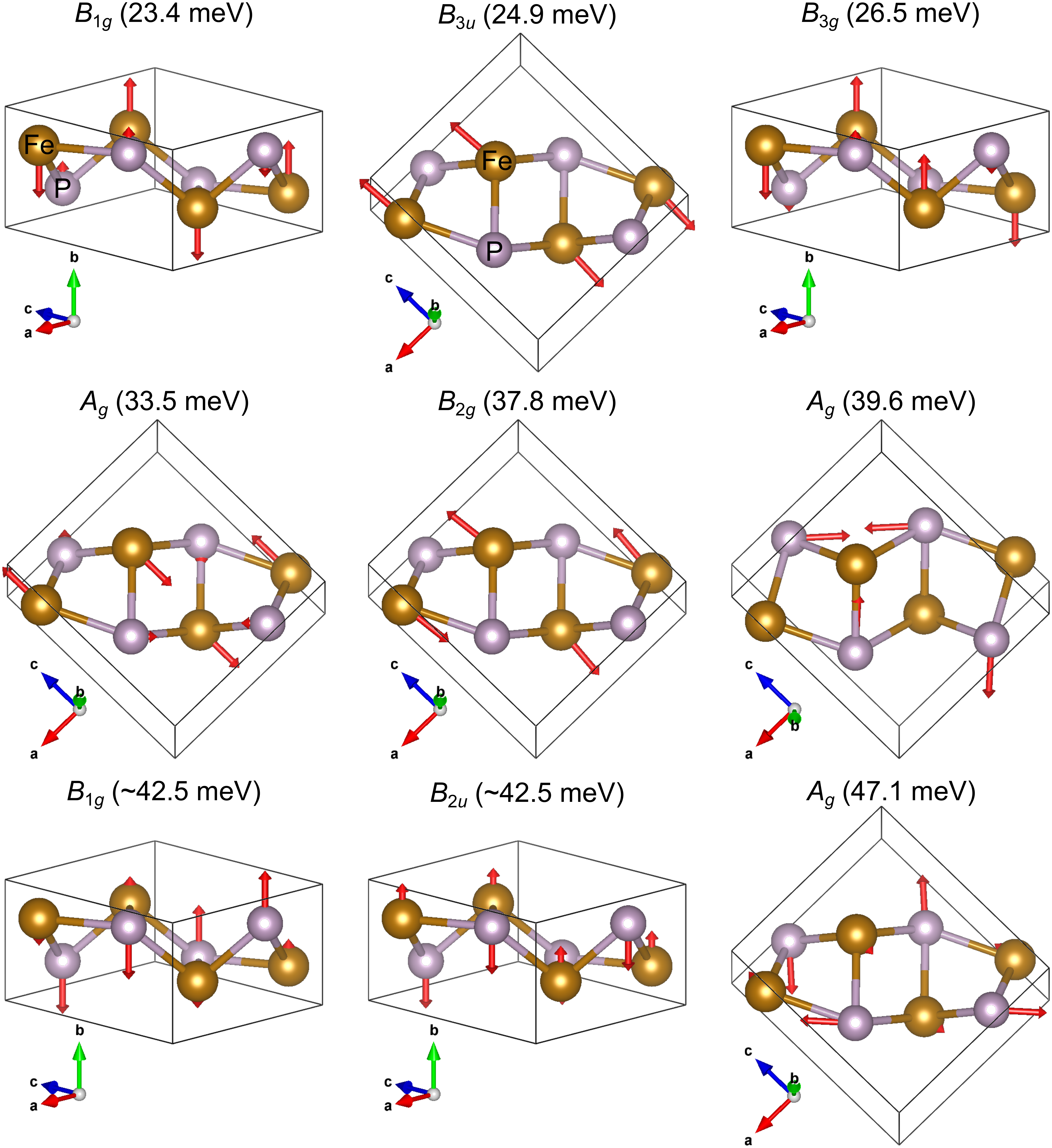}}
        \end{minipage}
        \caption{(color online). The displacement patterns of the optical phonons of Tab.~\ref{tab:tab1} in the center of the Brillouin zone. Arrows denote the eigenvectors. Large gold spheres are the Fe ions, small purple spheres are the P ions.}
        \label{ris:fig5}
\end{figure}

\begin{figure}[t]
        \begin{minipage}{0.99\linewidth}
        \center{\includegraphics[width=1\linewidth]{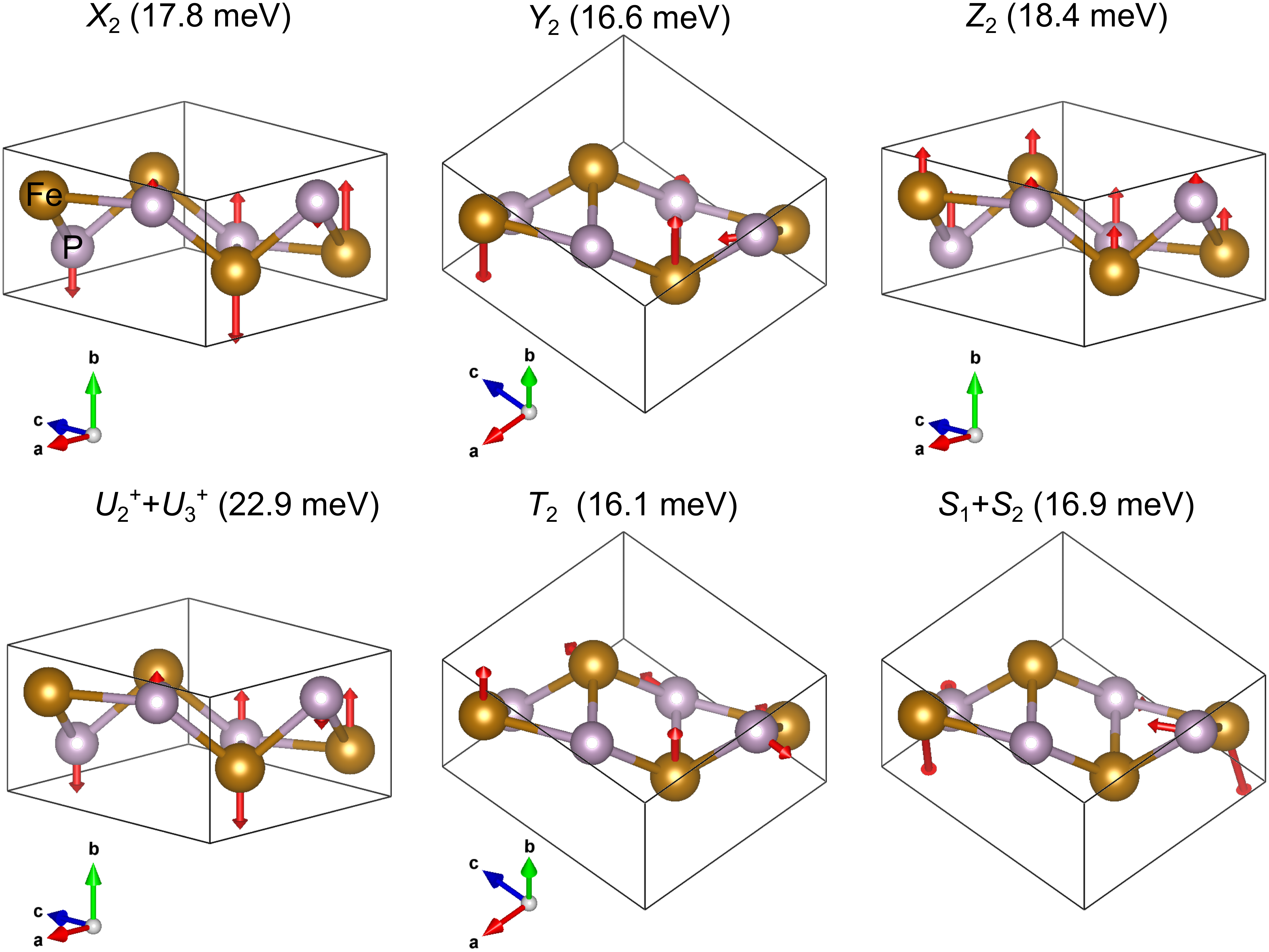}}
        \end{minipage}
        \caption{(color online). The displacement patterns of the lowest-energy phonons at the boundary of the Brillouin zone. Arrows denote the eigenvectors. Large gold spheres are the Fe ions, small purple spheres are the P ions.}
        \label{ris:fig6}
\end{figure}

\begin{figure*}
\includegraphics[width=\linewidth]{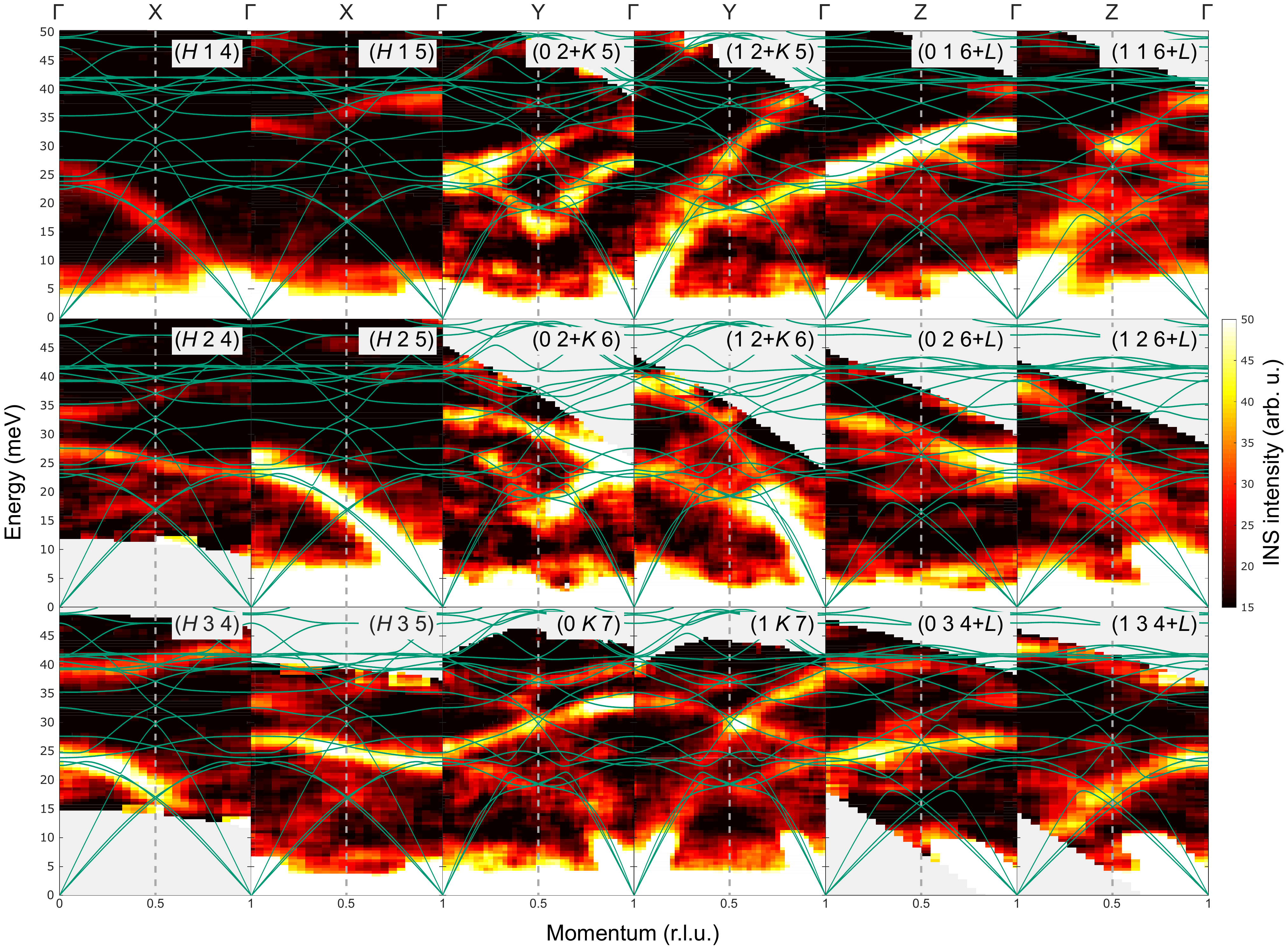}\vspace{3pt}
        \caption{(color online). The \textit{ab initio} phonon dispersion (solid lines) plotted over experimentally observed spectra (color-coded maps) along the $(H00)$, $(0K0)$, and $(00L)$ momentum directions corresponding to the $\Gamma$-$X$, $\Gamma$-$Y$, and $\Gamma$-$Z$ paths in the BZ. The data for each reduced momentum direction is shown by a set of six spectra at BZ adjacent to different $(HKL)$, where two of the three indeces alternate to demonstrate changes in the phonon spectral weight at different absolute momenta. Grey areas enclose $Q$-$E$ points where no data were collected.}
        \label{ris:fig7}
\end{figure*}

\begin{figure*}
\includegraphics[width=\linewidth]{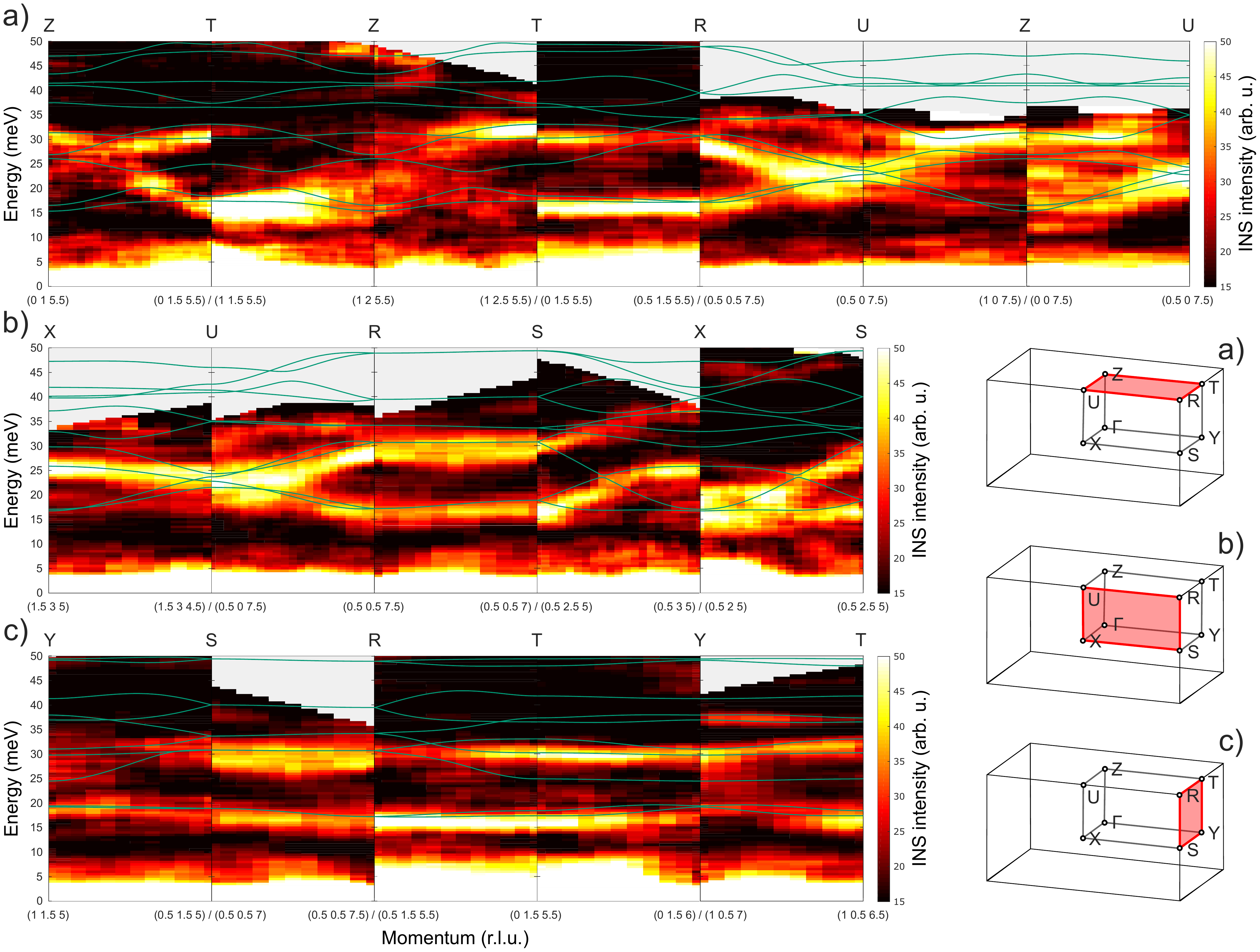}\vspace{3pt}
        \caption{(color online). The \textit{ab initio} phonon dispersion (solid lines) plotted over experimentally observed spectra (color-coded maps) along momentum paths connecting the high-symmetry points on the surface of the BZ (labelled on top) combined from the data at different total momentuma (marked at the bottom). (a)~The $Z$-$T$-$R$-$U$-$Z$ path. (b)~The $X$-$U$-$R$-$S$-$X$ path. (c)~The $Y$-$S$-$R$-$T$-$Y$ path. The surface area of the BZ enclosed by each loop is shaded in the diagrams for clarity.}
        \label{ris:fig8}
\end{figure*}

\begin{figure}[t]
        \begin{minipage}{0.99\linewidth}
        \center{\includegraphics[width=1\linewidth]{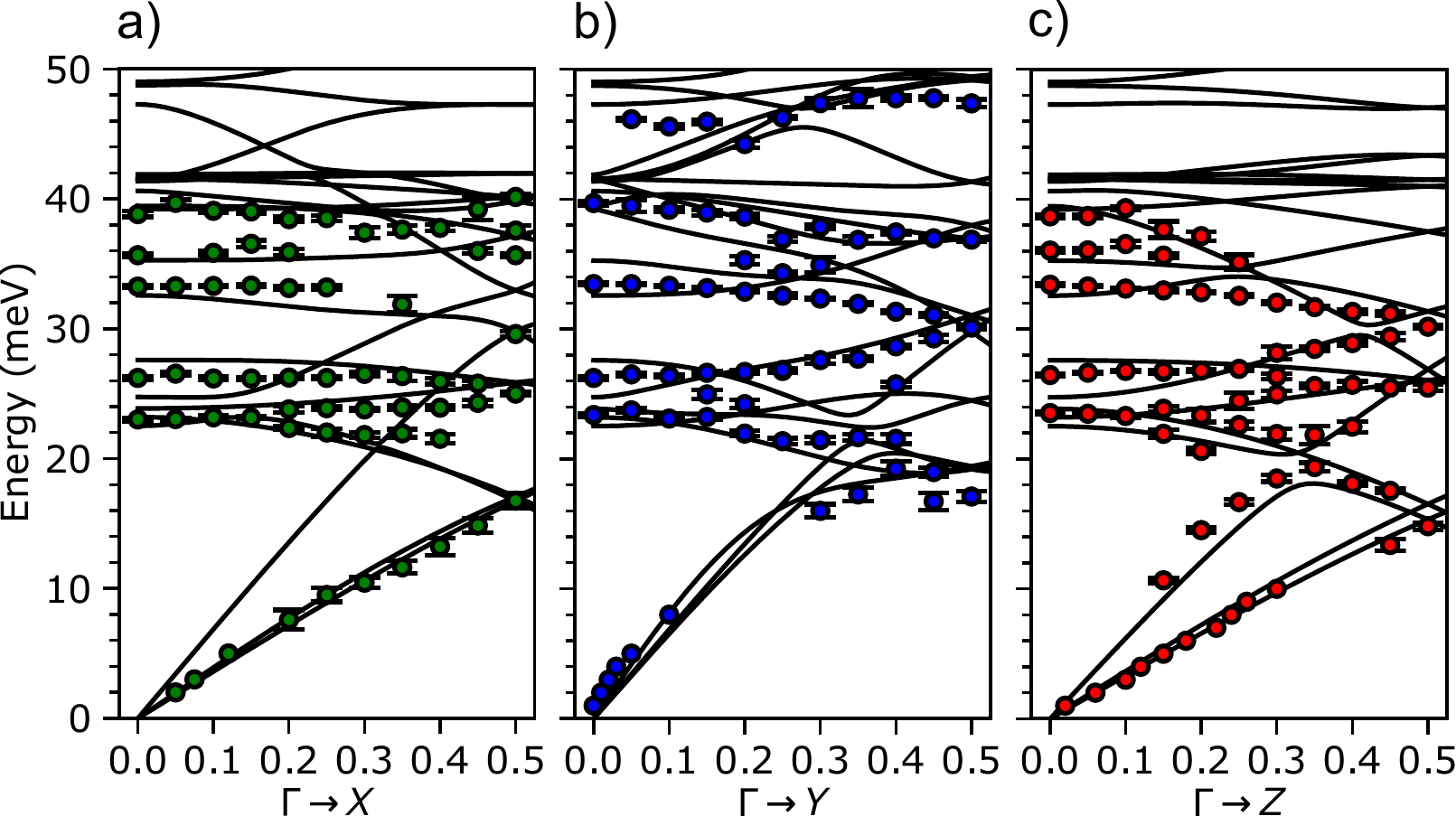}}
        \end{minipage}
        \caption{(color online). The results of the multi-zone fits for the $\Gamma$-$X$ (a), $\Gamma$-$Y$ (b), and $\Gamma$-$Z$ (c) paths of the BZ. Circles are experimental phonon energies extracted from the fits, solid lines are the DFT calculations.}
        \label{ris:fig9}
\end{figure}

We first discuss the INS spectra of FeP for momentum transfer along the [$00L$] (c*) direction. Figure~\ref{ris:fig1}(a) shows such a spectrum for energies $E$ up to 35 meV and momenta $\lvert \mathbf{Q} \rvert$ up to $\sim $10~\AA$^{-1}$. The spectrum reveals a clear sine-shaped mode that disperses within the $\sim$(25--35)~meV range and acquires a noticeable intensity at $\lvert \mathbf{Q} \rvert$ greater then $\sim$4~\AA$^{-1}$, such that two periods of the dispersion are included within the covered momentum transfer. The maxima of the dispersion coincide with $\lvert \mathbf{Q} \rvert \sim $4.32 and $\sim$6.48~\AA$^{-1}$, which corresponds to (004) and (006) reciprocal-lattice points, respectively, whereas the minima are found at the ($00L$) points with $L = 5$ and 7~r.l.u. A more detailed shape of the same mode is demonstrated in Fig.~\ref{ris:fig1}(b), where the data collected with a higher resolution are presented. The spectrum in Fig.~\ref{ris:fig1}(a) also features another mode with a much steeper dispersion emanating from the elastic ($E = 0$) line and reaching an energy of $\sim$27~meV at (00$\frac{15}{2}$). The first mode is identified as a longitudinal optical phonon branch. The second mode represents the longitudinal acoustic vibrations.

The elementary excitations of the helimagnet FeP at low temperatures consist of phonons and spin excitations, which can be separated by analysis of the $Q$-dependence of the intensity of a particular excitation. The intensity of the magnetic excitations is proportional to the magnetic form-factor, which drops rapidly with increasing $Q$, whereas the phonon scattering is proportional to $Q^2$ and leads to a larger cross-section for the increasing momentum transfer. Figs.~\ref{ris:fig1}(c) and \ref{ris:fig1}(d) show the scattering intensity as a function of $E$ for several fixed values of $Q$ [Fig.~\ref{ris:fig1}(c)] and as a function of $Q$ for the fixed energy $E = 30$~meV. At the (00$L$) reciprocal-lattice points, the observed INS peak intensity increases towards $L$. Figure~\ref{ris:fig1}(e) represents the integrated intensity of each peak extracted from the Gaussian fitting of the constant-$E$ intensity profile [Fig.~\ref{ris:fig1}(d)] plotted as a function of $Q$. As expected for scattering from phonons, the intensity follows the $I \propto Q^2$ trend drawn by the dashed line. Small deviations from the quadratic function are due to additional modulation by the structure factor, which alters the spectral weight of the optical phonon branch according to $L = 2n+1$ for even/odd values of $n$ at the (00$L$) point.

The observed spectral weight of each phonon branch at a specific reduced wave-vector $\mathbf{q}$ can significantly vary at different absolute momenta $\mathbf{Q}$. This enables optimizating the scattering geometry for the excitations of interest and performing mode assignments based on comparison of the measured and calculated intensities. Figures~\ref{ris:fig2}(a)--\ref{ris:fig2}(d) demonstrate the change of the INS intensity in the $E$-$\mathbf{Q}$ maps due to the phonon polarization factor, $I \propto \left( \mathbf{Q} \cdot \epsilon \right)^2$ ($\epsilon$ - the displacement vector), and the structure factor, $I \propto \left\vert \sum_d e^{i\mathbf{Q}\cdot\mathbf{r}_d}\right\vert^2$, of the multiatomic unit cell. In Fig.~\ref{ris:fig2}(a), the data were plotted along the closed path between the $\Gamma$-points of the BZs adjacent to ($HKL$) with $H = 1$ and $K$ and $L$ changing between 1--3, and 5--7 respectively, such that the momentum path reads: (125)--(136)--(127)--(116)--(125). The interval (3) is a replica of the interval (1) and features a steep acoustic mode dispersing up to $\sim$33~meV, whereas a mode with a bandwidth of $\sim$16~meV is observed at the intervals (2) and (4). Because the reduced wave-vector is almost parallel to the absolute momentum ($\mathbf{q} \parallel \mathbf{Q}$) along the (125)--(136) and (127)--(116) paths, the intensity of the longitudinal acoustic mode is maximized. The opposite holds for the momenta connecting (136) with (127) and (116) with (125), where the intensity of the transverse acoustic modes is much greater due to the $\mathbf{q} \perp \mathbf{Q}$ condition. If the same reciprocal-space loop is considered in the $H = 0$ plane [Fig.~\ref{ris:fig2}(c)], the intensity of the acoustic modes vanishes as the optical phonons acquire spectral weight due to the change of the phonon structure factor. The bottom optical branch (15--25~meV) mainly has the transverse character as its intensity is maximized for $\mathbf{q} \perp \mathbf{Q}$. The top mode, which disperses in the (25--35)~meV range, has a mixed character and carries similar spectral weight around the loop. The simulations of the INS intensity [Figs.~\ref{ris:fig2}(b) and \ref{ris:fig2}(d)] agree well with the experimental data.

\subsection{Spectra at the high-symmetry points}

It is essential to take into account the variation of the dynamic structure factor when the experimental data are compared with the results of the lattice-dynamics calculations. In order to resolve different modes in the experiment, it is also required to analyze the INS intensity in many different BZs. In Figs.~\ref{ris:fig3}(a)--\ref{ris:fig3}(d) and \ref{ris:fig4}(a)--\ref{ris:fig4}(a)--\ref{ris:fig4}(d), and in Table~\ref{tab:tab1}, we summarize the comparison between the experimental and calculated energies of the phonons at all eight high-symmetry points of the BZ. The constant-$\mathbf{Q}$ cuts through the data (the intensity as a function of energy) at the $\Gamma-$point are shown in Fig.~\ref{ris:fig3}(a). To extract the exact peak positions, the experimental data (symbols) were fitted by a sum of Gaussian functions (solid curve) with three free parameters per peak: the peak center, the peak width (FWHM), and the peak amplitude. The results of the simulations are plotted along with the data for a direct comparison (shaded area).

As can be seen, the theory predicts a strong peak at $\sim$35~meV and a weaker one at $\sim$40~meV at $\Gamma (017)$. According to their displacement vectors, these phonons correspond to the vibrations with the $A_g$ symmetry. Consistent with the predictions, the experimental profile shows two peaks with correct intensity ratio and positions. The first peak is at $E_{\text{exp}} = 33.6$~meV ($E_{\text{calc}} = 35.3$~meV) and the second is at $E_{\text{exp}} = 39.6$~meV ($E_{\text{calc}} = 41.5$~meV). The energies of both peaks are slightly lower than the calculated values, by 1.7 and 1.9~meV, respectively, which translates to the relative deviations, $\Delta E/E = \left(E_{\text{exp}}-E_{\text{calc}}\right)/E_{\text{exp}}$, of $-5.1$\% and $-4.8$\%. The spectrum at $\mathbf{Q} = \Gamma(016)$ exhibits a strong peak at $E \sim$25~meV that is in a good agreement with the simulated spectrum, which predicts only one peak at the zone center in the (016)~BZ. According to the calculated eigenvectors, the observed excitation can be identified as a $B_{3u}$ phonon. Overall, the profiles shown in Fig.~\ref{ris:fig3}(a) for seven different BZs demonstrate good agreement with the DFT results. Every INS spectrum at $\Gamma$ is characterized by dominant intensity of only a few modes, which can be well resolved in energy in different zones, except for the $\Gamma(134)$ spectrum where two modes have close energies and similar intensities merging into a single peak ($E_{\text{calc}} = 41.6$ and 41.9~meV) at $E_{\text{exp}} = 42.5$~meV. The complete list of the observed and calculated mode energies is given in Table~\ref{tab:tab1}.

Figures~\ref{ris:fig3}(b)--\ref{ris:fig3}(d) show examples of the spectra measured at the BZ-edge points $S$, $T$, and $U$, which correspond to the diagonal points of the rectangles formed by $\Gamma$-$X$ and $\Gamma$-$Y$, $\Gamma$-$Y$ and $\Gamma$-$Z$, and $\Gamma$-$X$ and $\Gamma$-$Z$, respectively. The spectra at $S(\frac{1}{2}\frac{1}{2}7)$ and $S(\frac{1}{2}\frac{5}{2}5)$ [Fig.~\ref{ris:fig3}(b)] feature two peaks at the same positions in both BZs, in agreement with the calculations. These peaks were identified as the vibrations with $S_1 + S_2$ symmetry observed at $\sim$17 and $\sim$30~meV.

Three modes are observed at $T(0\frac{3}{2}\frac{11}{2})$ and $T(1\frac{5}{2}\frac{11}{2})$. The low-lying mode with the $T_2$ symmetry and $E_{\text{exp}} = 16.1$~meV carries the spectral weight in both BZs, whereas the higher-energy modes with $T_1$ and $T_2$ symmetries and similar energies (30 and 32~meV) appear in different BZs. Four phonons are resolved at the $U$-point from the spectra of the ($\frac{3}{2}3\frac{9}{2}$) and ($\frac{1}{2}0\frac{15}{2}$) BZs [Fig.~\ref{ris:fig3}(d)].

It is worth noting that the lowest-energy phonon at $S$ and $T$ is softer than the calculated value by (8--12)\%. On the contrary, the relative disagreement between $E_{\text{calc}}$ and $E_{\text{exp}}$ for the vibrations that have higher energies at $S$ and $T$ amounts to only (1--7)\%. At the $U$-point, the low-lying mode is observed at the energy that is by 6\% higher than the predicted value. The softening (hardening) of the above mentioned modes at the corners of the BZ corresponds to the softening (hardening) of a transverse acoustic mode with momentum along $\Gamma$-$Y$ ($\Gamma$-$Z$).

The INS spectrum at the diagonal $R$-point is presented in Fig.~\ref{ris:fig4}(a). The $R$ and $S$-points are the only wavevectors at which the modes have four-fold degeneracy (there are six four-fold-degenerate modes at $R$ and $S$). All modes have two-fold degeneracy at all the other zone-edge high-symmetry points. The spectrum at $R(\frac{1}{2}\frac{3}{2}\frac{11}{2})$ reveals two excitations with $E_{\text{exp}} = 16$ and 29~meV. These are the lowest two four-fold-degenerate modes with $R_1 +R_2$ symmetry. The second mode closely matches the calculated energy ($E_{\text{calc}} = 30.7$~meV) but the first mode demonstrates a slightly larger discrepancy ($E_{\text{calc}} = 17.2$~meV). The mode softening is less pronounced than the softening of the lowest-energy mode at $S$ and $T$. This is expected as the $R$-points is further away from $Y$, where the largest mode softening (with respect to our DFT results) is observed.

The low-energy lattice vibrations at $X$ can be observed at a number of BZs (e.g. $X(\frac{1}{2}24)$, $X(\frac{1}{2}34)$, or $X(\frac{1}{2}25)$) as drawn in Fig.~\ref{ris:fig4}(b) where the highest INS intensity of these excitations (with $X_1$ and $X_2$ symmetry) is found at the (025) BZ center. The 17.8~meV peak at $X(\frac{1}{2}25)$ represents a mixture of two modes with close energies. Similarly, two phonons with energies of 37.2 and 39.8~meV are predicted to have spectral weight at $X(\frac{1}{2}24)$, where they are not resolved and seen as a single 37.1-meV peak.

The intensity profile of $Y(0\frac{5}{2}5)$ [Fig.~\ref{ris:fig4}(c)] reveals a transverse acoustic mode along the $\Gamma$-$Y$ path (the $(0K0)$ direction) that exhibits large softening compared with the DFT results. The experimental energy $E_{\text{exp}}=16.6$~meV deviates by 2.6~meV from the simulated value ($E_{\text{calc}} = 19.2$~meV). The other modes at $Y$ are in a better agreement with the theory. The transverse acoustic mode at $Z$ reaches 18.4~meV, which is 1.9~meV larger than $E_{\text{calc}}$. (See the spectrum at $Z(13\frac{9}{2}) $ in Fig.~\ref{ris:fig4}(d)). A number of the other modes were observed at the $Z$-point gthat are in a good agreement with the $E_{\text{calc}}$ values.

The values of $E_{\text{exp}}$ and $E_{\text{calc}}$ for all the experimentally determined vibrations at the high-symmetry points of the BZ are listed in Table~\ref{tab:tab1}. If the relative discrepancies $\Delta E/E$ of all the observed modes are considered together, it can be found that the average $\Delta E/E$ amounts to $\sim$ $-3$\% and its standard deviation is $\sim$4.5\%. This indicates that our DFT calculations tend to slightly overestimate the phonon energies. If the calculated dispersions are renormalized by a factor of 0.97, the agreement between $E_{\text{exp}}$ and $E_{\text{calc}}$ are found within $\pm 5$\% for all the modes, except for the transverse mode at $T$, $S$, and $Y$ (softening, $\Delta E/E = 5$\%--13\%, from $T$ to $Y$), and at $X$, $U$, and $Z$ (hardening, $\Delta E/E = 7$\%--13\%, from $X$ to $Z$).

\subsection{Displacement patterns}

In the previous section we have shown that our calculations quantitatively reproduce the dispersion and intensity distribution of the observed phonon spectrum at all high symmetry points of the BZ. This allowed us to identify the character of the lattice vibrations by analyzing the calculated eigenvectors of the observed phonons. Figures~\ref{ris:fig5} and~\ref{ris:fig6} show selected sketches of the lattice vibrations resulting from the calculated eigenvectors at the $\Gamma$- and other high-symmetry points of the BZ, respectively. Note that at the $\Gamma$-point we present only those phonons which were identified from the experiment [see Tab.~\ref{tab:tab1}]. Fig.~\ref{ris:fig6} shows the lowest-energy phonons for each high-symmetry point at the BZ boundary.

The vibrations that have energies $E < 40$~meV mainly involve Fe ions, whereas P ions contribute mostly to the higher-energy phonons. These results agree well with naive expectations, because Fe atoms are approximately twice as heavy as the P atoms. It is worth noting that the phonons depicted in Figs.~\ref{ris:fig5} and \ref{ris:fig6} mainly involve atomic displacement orthogonal to the $[100]$ crystal direction. That is because our experimental geometry we had limited access to the $(H00)$ direction of the reciprocal space. Therefore, we could hardly detect the phonons with polarization (displacement vectors) parallel to the $a$ axis due to the polarization factor $\left( \mathbf{Q} \cdot \epsilon \right)^2$ in the INS cross-section Eq.~\eqref{eq:eq1}.

It is convenient to consider the displacements of Fe ions of the first five vibrations of Fig.~\ref{ris:fig5} with respect to the quasi-hexagonal crystal structure (see Fig.~\ref{ris:fig0}(c)). The 23.4~meV excitations ($B_{1g}$) involve the displacements of the Fe ions along the $b$-axis, such that the three neighbouring sites within the triangular layers cause the triangles to rotate around the axis perpendicular to the layers (the $a$-axis). The optical phonons of 26.5 and 33.5~meV represent the parallel translation of the adjacent triangular layers (in the opposite direction) along $b$ and $c$, respectively. The $B_{3u}$ and $B_{2g}$ vibrations (24.9~meV and 37.8~meV) correspond to a contraction/elongation of the triangular layers along $b$. The 24.9~meV phonon recovers the hexagonal symmetry for the one Fe layer, but leads to a larger distortion of the other layer, whereas the 37.8-meV excitation leads to the symmetry recovery for both layers (the elongated layer is contracted and the contracted layer is elongated along $b$). The $A_g$ vibrations (39.6 and 47.1~meV) involve a complex displacement of P ions in the $ac$ plane, and the phonons of 42.5~meV are characterized by a shift of P ions along $b$, which corresponds to a parallel translations of the P triangular layers along the $b$-axis (either opposite to the adjacent layers or in the same direction).

The lowest-energy phonons at the BZ boundary can be described as follows. The excitations at the symmetry points $X$, $Z$, and $U$ correspond to the displacements only along $b$ (of both Fe and P ions), whereas the vibrations at $Y$, $T$, and $S$ have complex displacement patterns , where different sites move along different crystallographic directions. The 16.6~meV phonon of the $Y$-point ($Y_2$ symmetry) is of particular interest as this is the excitations that shows softening as compared to the results of our DFT calculations. It involves a larger displacement of two Fe ions along $b$, and a smaller displacement of two P ions in the $ac$ plane. The movement of the Fe sublattice corresponds to the rotation of triangles within the layers where the adjacent layers rotate in the same direction.

\subsection{Phonon dispersions}

Having discussed the phonon modes at the high symmetry points of the BZ, we turn to following the phonon dispersion along the high symmetry directions. Figure~\ref{ris:fig7} summarizes the experimental data for the $\Gamma$-$X$, $\Gamma$-$Y$, and $\Gamma$-$Z$ paths in the BZ, which coincide with the ($H00$), $(0K0)$, and $(00L)$ directions, respectively. Because the INS spectral weight for different modes varies drastically as reciprocal space is traversed, one has to consider many different BZs, as shown in Fig.~\ref{ris:fig7}.

Only the 24.9~meV mode bears a significant spectral weight at $\Gamma(014)$. At small reduced momenta along ($H14$), the spectral weight is transferred to the two lowest energy optical modes, which disperse downwards and reach the energies of $\sim$17.8~meV at the $X$-point at the BZ boundary. At the momenta between $(\frac{1}{2}14)$ and $(114)$, the entire observed spectral weight transfers to the TA modes, which disperse in a good agreement with our \textit{ab initio} calculations. As can be seen in Fig.~\ref{ris:fig7}, the spectral weight is altered when the momentum is changed from $L = 4$ to $L = 5$ for the same $H$ and $K$. In the (015) BZ, the $A^2_g$ mode has the highest intensity. It weakly disperses downwards and fades out on approach to the zone boundary. The spectral weight then transfers to the weakly dispersing $B^1_{2g}$ mode in the ($115$) zone.

The INS intensity follows the structure factor. Thus, the same modes that are seen along the ($H14$) for $0 < H < 1$ can be observed in $(H25)$ and $(H34)$, which yields the rule $K, L \pm 1$ for a given $H$ [Fig.~\ref{ris:fig7}]. In addition, the polarization factor makes the 39.6~meV mode visible, as it has a longitudinal polarization character. Further, the dispersion of the $B^2_{1g}$ and $B^1_{2u}$ modes can be followed in the ($134$) BZ. The dispersion relation of the $B^1_{3g}$ and $B^1_{1g}$ modes can be followed in the $(H24)$ and $(H35)$ BZs for $0 < H < 1$, where they carry the maximal spectral weight. No spectral weight is present for the acoustic mode at these zones due to the structure factor.

The acoustic phonon intensity obeys the following systematic absence rules for the dispersion along the $\Gamma$-$Y$ path, as we show in Fig.~\ref{ris:fig7}. The acoustic modes have nonzero intensity at the $(125)$, $(136)$, and $(107)$, which is $K + L$ odd for $H = 1$, whereas only the optical modes are observed in the $(0KL)$ zones. The same applies for the acoustic dispersions along $\Gamma$-$Z$.

As mentioned in Sec.~\ref{Sec:III}B, the $Y_2$ mode shows a notable softening compared to the prediction of our \textit{ab initio} calculations. The dispersion of this mode in the vicinity of the $Y$-point is clearly observed in the $(0~2+K~5)$ and $(0~2+K~6)$ zones for $0< K < 1$, where it shows a much steeper upward dispersion than predicted by the calculations. On the contrary, the $Y_1$ mode has a shallower dispersion in perfect agreement with the theory, as can be seen in the vicinity of the points $(1\frac{5}{2}5)$, $(1\frac{5}{2}6$), and $(1\frac{1}{2}7)$ in Fig.~\ref{ris:fig7}. The $\Gamma$-$Z$ dispersion of the LA mode is characterized by large intensity in the $(116)$ zone due to the polarization factor. Weaker intensities of both the LA and TA modes are observed at $(134)$. The calculated energies and dispersions of the optical phonons are in good agreement with the INS data along the $\Gamma$-$Z$ path as evidenced in Fig.~\ref{ris:fig7}.

We also resolved the phonon dispersions at momenta connecting all the high-symmetry points of the BZ boundary. Figure~\ref{ris:fig8}(a) shows a cut through the INS data along the momentum path $Z$-$T$-$R$-$U$-$Z$. To account for the spectral weight alteration in different BZs, we combined the spectra at the $(015)$ and $(125)$ BZs for $Z$-$T$ and the $(107)$ and $(007)$ BZs for $U$-$Z$. The INS spectra in Fig.~\ref{ris:fig8}(a) cover the first five modes dispersing between $\sim$15 and $\sim$40~meV. All observed modes exhibit strong dispersion except along the $T$-$R$ path, where a flat 16 meV mode is observed. This is in agreement with the lowest energy mode predicted by the DFT calculations. The data for the $X$-$U$-$R$-$S$-$X$ and $Y$-$S$-$R$-$T$-$Y$ paths are compared with the calculated dispersions in Figs.~\ref{ris:fig8}(b) and \ref{ris:fig8}(c), respectively, which also demonstrate a relatively good agreement for all momenta except for the lowest-lying mode along $T$-$Y$. As can be seen from Fig.~\ref{ris:fig8}(c), the $Y_2$ mode softens (as compared to the DFT predictions) not only along $\Gamma$-$Y$ (Fig.~\ref{ris:fig7}), but also along $T$-$Y$, where it is flat. Meanwhile, the closely-spaced $Y_1$ modes are in good agreement with the theoretically-predicted dispersion along $T$-$Y$. All the modes retain the four-fold degeneracy along $R$-$S$ as a result of a pair of the nonsymmorphic symmetries of the $Pnma$ space group~\cite{PhysRevMaterials.3.095004}.

To compare the observed optical phonon energies with the results of our first-principles calculations in detail, we performed multi-zone fitting uning the entire unsymmetrized INS dataset~\cite{PhysRevB.89.064310}. In multi-zone fitting, the energy and line-width of each mode at a given reduced momentum is extracted from the experimental data in many BZs simultaneously (for the same reduced momentum). This allows one to significantly improve statistics and to resolve modes that are close in energy but whose spectral weights differ in many BZs. During the multi-zone fitting, the intensity of each peak is allowed to vary in every BZ as each mode has different spectral weight in different BZs. However, the positions and widths of the peaks are fixed to be the same in every zone and refined globally. As a result, the extracted peak positions from all the BZs for the same crystal momentum direction can be plotted together, as shown in Figs.~\ref{ris:fig9}(a)--\ref{ris:fig9}(c) for the $\Gamma$-$X$, $\Gamma$-$Y$, and $\Gamma$-$Z$ paths, respectively. To perform the multi-zone fitting, we used the \textsc{phonon explorer} software~\cite{https://github.com/dmitryr1234/phonon-explorer,Ahmadova2020} to fit constant-momentum cuts assuming Gaussian peak shapes for every phonon. Linear background was subtracted from every constant-momentum cut before the multi-zone fit was performed.

The positions of the acoustic modes at low energies below 10~meV were extracted from constant-energy cuts. For this, the $I(Q)$ profiles were fitted with a Gaussian function for every fixed energy.

As can be seen in Figs.~\ref{ris:fig9}(a), a good agreement between the DFT results and experimental data is found for all resolved modes for $\Gamma$-$X$. The slope of the TA modes matches the predicted energies for all momenta from the center of the zone to the zone boundary. However, due to a finite coverage of the $(H00)$ direction in our experimental setup, it is not possible to verify the DFT predictions of the dispersion of the LA mode for $\Gamma$-$X$. A good agreement is also seen for many weakly dispersing optical phonon branches in the 20--40~meV range.

Because the LA and TA modes along the $\Gamma$-$Y$ path are close in energy, we fit the constant-momentum INS data for $E < 10$~meV with only one peak function. Figure~\ref{ris:fig9}(b) shows that the experimental energies of the acoustic modes for the reduced momentum below 0.1~r.l.u. agree well with the DFT predictions. As was discussed in relation to Fig.~\ref{ris:fig7}, the $Y_2$ mode exhibits a significant softening of its dispersion in the vicinity of the $Y$-point. The softening of this low-energy mode as compared to the DFT results is clear in Fig.~\ref{ris:fig9}(b). The discrepancy starts at $\sim$0.3~r.l.u. and increases as the momentum approaches the $Y$-point at 0.5~r.l.u. The other optical modes demonstrate a good agreement at all momenta.

Figure~\ref{ris:fig9}(c) summarizes the extracted peak positions for $\Gamma$-$Z$. In this reciprocal-space direction, the TA and LA modes have significantly different slopes and could be resolved seperately in our INS data. The experimental and theoretical dispersions of the LA mode follow the same trend, but the experimentally obtained mode has slightly higher energies for all the momenta. The dispersions of the TA branches are found to be in good agreement. Similarly to $\Gamma$-$X$ and $\Gamma$-$Y$, the optical dispersions for $\Gamma$-$Z$ show good agreement between the DFT results and the INS data.

\section{Discussion and Conclusions}\label{Sec:IV}

Since the magnetic structure of FeP is noncollinear, a significant renormalization of its magnon and phonon spectra can be expected in the presence of magnetoelastic coupling. A linear magnon-phonon term, which is absent in the case of collinear spin order, becomes nonzero in the presence of a noncollinear magnetic structure. This is a consequence of the absence of a global spin-quantization axis which allows for a one-magnon term in the spin Hamiltonian~\cite{toth2016electromagnon,oh2016spontaneous}. This also means magnon-phonon hybridization causes stronger INS spectral renormalization than the magnon-magnon and the phonon-phonon interactions~\cite{PhysRevB.97.201113,PhysRevB.94.104421,PhysRevB.99.214445}.

The present comprehensive investigation of the lattice dynamics shows that the spin-lattice coupling is seemingly not playing a major role in FeP. This is evidenced by a very good overall agreement between the observed phonon spectra and the DFT calculations that do not take into account the magnetoelastic coupling. The ordering temperature of FeP yields an approximate expected magnon bandwidth of an order of a few dozen meV, or more, if a significant frustration of the exchange interactions is present. An estimate of the magnon bandwidth in FeP can also be made by comparison with CrAs, which is isostructural and orders into the same double-helix magnetic structure. Recent INS measurement of a CrAs powder sample showed that the magnetic excitations reach at least $\sim$100~meV. Taking into account that $T_{\text N}$ of CrAs is twice higher than that of FeP, an estimate of a 50~meV magnon bandwidth can be made for the latter. Since the influence of the magnon-phonon coupling is typically stronger in the vicinity of the crossing point of the bare modes~\cite{PhysRevB.99.214445}, it is unlikely that the observed softening of the $Y_2$ phonon is caused by the overlapping magnon branches as those are expected at a higher energy at the BZ boundary. However, detailed INS measurements of the spin excitations of FeP are required to exclude the magnetoelastic origin of the softening of the $Y_2$ mode, which will be addressed in future studies.

Measurements of the temperature dependence of the phonon energies in the vicinity of $T_{\text N}$ might also provide useful information on the magnon-phonon coupling in this compound. For example, the lattice vibrations of the isostructural CrAs at the zone center were recently studied by Raman scattering~\cite{PhysRevB.100.104301}. Sen \textit{et al.}~\cite{PhysRevB.100.104301} resolved all four full-symmetric $A_g$ modes with the energies of $\sim$14, 23, 28, and 32~meV. The energies of the $A_g$ modes of FeP determined in our study ($\sim$22--47~meV) are higher due to a lighter mass of the P ions, in agreement with the theoretical expectations. It was shown that the lowest-energy $A_g$ mode in CrAs undergoes a large ($\sim$12\%) renormalization at $T_{\text N}$. It is therefore very intriguing to address the energy shift and the broadening of the phonon modes in FeP across the magnetic ordering temperature in future studies, for which the full characterization of the phonons presented in our work is indispensable. The detailed phonon analysis of FeP should serve as a foundation for further studies of the lattice dynamics of other representatives of the 3$d$-metal monopnictide family, where a more complex interplay between the charge, orbital, spin, and lattice degrees of freedom might be present.

The overall success of the DFT calculations to capture the lattice dynamics in FeP, as presented in our study, demonstrates that the DFT method may be applicable for investigations of the pressure-induced phenomena in related materials, such as the strong anisotropic compression and isostructural transitions in MnP~\cite{Han2018} and CrAs~\cite{Yu14766}. This may allow one to address the important problem of the pressure-induced superconductivity in these compounds.

To conclude, we have performed INS measurements of lattice dynamics in FeP. The phonon spectra were collected in a large part of reciprocal space for energies up to 50~meV, which allowed us to resolve the majority of the acoustic and optical modes. We discussed momentum-dependence of the phonon spectral weight across many BZs and demonstrated the role of the nuclear structure and polarization factors in the variation of the observed INS intensities. With the help of accurate \textit{ab initio} calculations, we were able to assign specific symmetries to the observed excitations and determine the displacement patterns for all modes. A detailed comparison between the theoretical phonon energies and the experimental data was considered at all high-symmetry points in the BZ, as well as all momenta along the high-symmetry directions in the entire BZ. The experimentally resolved phonon dispersions and the observed intensities of the specific modes showed a good agreement with the DFT predictions, except for the lowest-energy mode at the $Y$-point of the BZ boundary, which exhibit a noticeable softening of $\sim$13\% as compared to the calculations. The origin of this discrepancy will be addressed in future studies.

\section*{Acknowledgments}

I.V.M. thanks RSF 19-43-04129 for financial support in the frame of the joint DFG-RSF project, I.C. thanks the RFBR fund (grant 18-33-01282 https://kias.rfbr.ru/index.php). D.S.I. acknowledges support from the German Research Foundation (DFG) under Grant No. IN 209/9-1, via the project C03 of the Collaborative Research Center SFB 1143 (project-id 247310070) at the TU Dresden and the W\"{u}rzburg-Dresden Cluster of Excellence on Complexity and Topology in Quantum Matterials\,---\,\textit{ct.qmat} (EXC 2147, project-id 390858490). S.A. acknowledges DFG funding No. AS 523$\backslash$4-1. This work used Phonon Explorer software for data analysis. Work at the University of Colorado-Boulder was supported by the U.S. Department of Energy, Office of Basic Energy Sciences, Office of Science, under Contract No. DE-SC0006939.

\bibliography{FeP}

\begin{thebibliography}{54}%
\makeatletter
\providecommand \@ifxundefined [1]{%
 \@ifx{#1\undefined}
}%
\providecommand \@ifnum [1]{%
 \ifnum #1\expandafter \@firstoftwo
 \else \expandafter \@secondoftwo
 \fi
}%
\providecommand \@ifx [1]{%
 \ifx #1\expandafter \@firstoftwo
 \else \expandafter \@secondoftwo
 \fi
}%
\providecommand \natexlab [1]{#1}%
\providecommand \enquote  [1]{``#1''}%
\providecommand \bibnamefont  [1]{#1}%
\providecommand \bibfnamefont [1]{#1}%
\providecommand \citenamefont [1]{#1}%
\providecommand \href@noop [0]{\@secondoftwo}%
\providecommand \href [0]{\begingroup \@sanitize@url \@href}%
\providecommand \@href[1]{\@@startlink{#1}\@@href}%
\providecommand \@@href[1]{\endgroup#1\@@endlink}%
\providecommand \@sanitize@url [0]{\catcode `\\12\catcode `\$12\catcode
  `\&12\catcode `\#12\catcode `\^12\catcode `\_12\catcode `\%12\relax}%
\providecommand \@@startlink[1]{}%
\providecommand \@@endlink[0]{}%
\providecommand \url  [0]{\begingroup\@sanitize@url \@url }%
\providecommand \@url [1]{\endgroup\@href {#1}{\urlprefix }}%
\providecommand \urlprefix  [0]{URL }%
\providecommand \Eprint [0]{\href }%
\providecommand \doibase [0]{http://dx.doi.org/}%
\providecommand \selectlanguage [0]{\@gobble}%
\providecommand \bibinfo  [0]{\@secondoftwo}%
\providecommand \bibfield  [0]{\@secondoftwo}%
\providecommand \translation [1]{[#1]}%
\providecommand \BibitemOpen [0]{}%
\providecommand \bibitemStop [0]{}%
\providecommand \bibitemNoStop [0]{.\EOS\space}%
\providecommand \EOS [0]{\spacefactor3000\relax}%
\providecommand \BibitemShut  [1]{\csname bibitem#1\endcsname}%
\let\auto@bib@innerbib\@empty
\bibitem [{\citenamefont {Felcher}\ \emph {et~al.}(1971)\citenamefont
  {Felcher}, \citenamefont {Smith}, \citenamefont {Bellavance},\ and\
  \citenamefont {Wold}}]{felcher1971magnetic}%
  \BibitemOpen
  \bibfield  {author} {\bibinfo {author} {\bibfnamefont {G.}~\bibnamefont
  {Felcher}}, \bibinfo {author} {\bibfnamefont {F.}~\bibnamefont {Smith}},
  \bibinfo {author} {\bibfnamefont {D.}~\bibnamefont {Bellavance}}, \ and\
  \bibinfo {author} {\bibfnamefont {A.}~\bibnamefont {Wold}},\ }\href {\doibase
  10.1103/PhysRevB.3.3046} {\bibfield  {journal} {\bibinfo  {journal} {Phys.
  Rev. B}\ }\textbf {\bibinfo {volume} {3}},\ \bibinfo {pages} {3046} (\bibinfo
  {year} {1971})}\BibitemShut {NoStop}%
\bibitem [{\citenamefont {H{\"a}ggstr{\"o}m}\ and\ \citenamefont
  {Narayanasamy}(1982)}]{haggstrom1982mossbauer}%
  \BibitemOpen
  \bibfield  {author} {\bibinfo {author} {\bibfnamefont {L.}~\bibnamefont
  {H{\"a}ggstr{\"o}m}}\ and\ \bibinfo {author} {\bibfnamefont {A.}~\bibnamefont
  {Narayanasamy}},\ }\href {\doibase 10.1016/0304-8853(82)90205-0} {\bibfield
  {journal} {\bibinfo  {journal} {J. Magn. Magn. Mater.}\ }\textbf {\bibinfo
  {volume} {30}},\ \bibinfo {pages} {249} (\bibinfo {year} {1982})}\BibitemShut
  {NoStop}%
\bibitem [{\citenamefont {Felcher}(1966)}]{felcher1966magnetic}%
  \BibitemOpen
  \bibfield  {author} {\bibinfo {author} {\bibfnamefont {G.~P.}\ \bibnamefont
  {Felcher}},\ }\href {\doibase 10.1063/1.1708333} {\bibfield  {journal}
  {\bibinfo  {journal} {J. Appl. Phys.}\ }\textbf {\bibinfo {volume} {37}},\
  \bibinfo {pages} {1056} (\bibinfo {year} {1966})}\BibitemShut {NoStop}%
\bibitem [{\citenamefont {Pan}\ \emph {et~al.}(2019)\citenamefont {Pan},
  \citenamefont {Jang}, \citenamefont {Lee}, \citenamefont {Sutarto},
  \citenamefont {He}, \citenamefont {Zeng}, \citenamefont {Liu}, \citenamefont
  {Zhang}, \citenamefont {Feng}, \citenamefont {Hao} \emph
  {et~al.}}]{pan2019intertwined}%
  \BibitemOpen
  \bibfield  {author} {\bibinfo {author} {\bibfnamefont {B.}~\bibnamefont
  {Pan}}, \bibinfo {author} {\bibfnamefont {H.}~\bibnamefont {Jang}}, \bibinfo
  {author} {\bibfnamefont {J.-S.}\ \bibnamefont {Lee}}, \bibinfo {author}
  {\bibfnamefont {R.}~\bibnamefont {Sutarto}}, \bibinfo {author} {\bibfnamefont
  {F.}~\bibnamefont {He}}, \bibinfo {author} {\bibfnamefont {J.}~\bibnamefont
  {Zeng}}, \bibinfo {author} {\bibfnamefont {Y.}~\bibnamefont {Liu}}, \bibinfo
  {author} {\bibfnamefont {X.}~\bibnamefont {Zhang}}, \bibinfo {author}
  {\bibfnamefont {Y.}~\bibnamefont {Feng}}, \bibinfo {author} {\bibfnamefont
  {Y.}~\bibnamefont {Hao}},  \emph {et~al.},\ }\href {\doibase
  10.1103/PhysRevX.9.021055} {\bibfield  {journal} {\bibinfo  {journal} {Phys.
  Rev. X}\ }\textbf {\bibinfo {volume} {9}},\ \bibinfo {pages} {021055}
  (\bibinfo {year} {2019})}\BibitemShut {NoStop}%
\bibitem [{\citenamefont {Yang}\ \emph {et~al.}(2010)\citenamefont {Yang},
  \citenamefont {Xie}, \citenamefont {Zhang}, \citenamefont {He}, \citenamefont
  {Ge}, \citenamefont {Wang}, \citenamefont {Chen}, \citenamefont {Arita},
  \citenamefont {Jiang}, \citenamefont {Shimada}, \citenamefont {Taniguchi},
  \citenamefont {Vobornik}, \citenamefont {Rossi}, \citenamefont {Hu},
  \citenamefont {Lu}, \citenamefont {Shen}, \citenamefont {Lu},\ and\
  \citenamefont {Feng}}]{PhysRevB.82.104519}%
  \BibitemOpen
  \bibfield  {author} {\bibinfo {author} {\bibfnamefont {L.~X.}\ \bibnamefont
  {Yang}}, \bibinfo {author} {\bibfnamefont {B.~P.}\ \bibnamefont {Xie}},
  \bibinfo {author} {\bibfnamefont {Y.}~\bibnamefont {Zhang}}, \bibinfo
  {author} {\bibfnamefont {C.}~\bibnamefont {He}}, \bibinfo {author}
  {\bibfnamefont {Q.~Q.}\ \bibnamefont {Ge}}, \bibinfo {author} {\bibfnamefont
  {X.~F.}\ \bibnamefont {Wang}}, \bibinfo {author} {\bibfnamefont {X.~H.}\
  \bibnamefont {Chen}}, \bibinfo {author} {\bibfnamefont {M.}~\bibnamefont
  {Arita}}, \bibinfo {author} {\bibfnamefont {J.}~\bibnamefont {Jiang}},
  \bibinfo {author} {\bibfnamefont {K.}~\bibnamefont {Shimada}}, \bibinfo
  {author} {\bibfnamefont {M.}~\bibnamefont {Taniguchi}}, \bibinfo {author}
  {\bibfnamefont {I.}~\bibnamefont {Vobornik}}, \bibinfo {author}
  {\bibfnamefont {G.}~\bibnamefont {Rossi}}, \bibinfo {author} {\bibfnamefont
  {J.~P.}\ \bibnamefont {Hu}}, \bibinfo {author} {\bibfnamefont {D.~H.}\
  \bibnamefont {Lu}}, \bibinfo {author} {\bibfnamefont {Z.~X.}\ \bibnamefont
  {Shen}}, \bibinfo {author} {\bibfnamefont {Z.~Y.}\ \bibnamefont {Lu}}, \ and\
  \bibinfo {author} {\bibfnamefont {D.~L.}\ \bibnamefont {Feng}},\ }\href
  {\doibase 10.1103/PhysRevB.82.104519} {\bibfield  {journal} {\bibinfo
  {journal} {Phys. Rev. B}\ }\textbf {\bibinfo {volume} {82}},\ \bibinfo
  {pages} {104519} (\bibinfo {year} {2010})}\BibitemShut {NoStop}%
\bibitem [{\citenamefont {Chu}\ \emph {et~al.}(2010)\citenamefont {Chu},
  \citenamefont {Analytis}, \citenamefont {De~Greve}, \citenamefont {McMahon},
  \citenamefont {Islam}, \citenamefont {Yamamoto},\ and\ \citenamefont
  {Fisher}}]{Chu824}%
  \BibitemOpen
  \bibfield  {author} {\bibinfo {author} {\bibfnamefont {J.-H.}\ \bibnamefont
  {Chu}}, \bibinfo {author} {\bibfnamefont {J.~G.}\ \bibnamefont {Analytis}},
  \bibinfo {author} {\bibfnamefont {K.}~\bibnamefont {De~Greve}}, \bibinfo
  {author} {\bibfnamefont {P.~L.}\ \bibnamefont {McMahon}}, \bibinfo {author}
  {\bibfnamefont {Z.}~\bibnamefont {Islam}}, \bibinfo {author} {\bibfnamefont
  {Y.}~\bibnamefont {Yamamoto}}, \ and\ \bibinfo {author} {\bibfnamefont
  {I.~R.}\ \bibnamefont {Fisher}},\ }\href {\doibase 10.1126/science.1190482}
  {\bibfield  {journal} {\bibinfo  {journal} {Science}\ }\textbf {\bibinfo
  {volume} {329}},\ \bibinfo {pages} {824} (\bibinfo {year}
  {2010})}\BibitemShut {NoStop}%
\bibitem [{\citenamefont {Yi}\ \emph {et~al.}(2011)\citenamefont {Yi},
  \citenamefont {Lu}, \citenamefont {Chu}, \citenamefont {Analytis},
  \citenamefont {Sorini}, \citenamefont {Kemper}, \citenamefont {Moritz},
  \citenamefont {Mo}, \citenamefont {Moore}, \citenamefont {Hashimoto},
  \citenamefont {Lee}, \citenamefont {Hussain}, \citenamefont {Devereaux},
  \citenamefont {Fisher},\ and\ \citenamefont {Shen}}]{Yi6878}%
  \BibitemOpen
  \bibfield  {author} {\bibinfo {author} {\bibfnamefont {M.}~\bibnamefont
  {Yi}}, \bibinfo {author} {\bibfnamefont {D.}~\bibnamefont {Lu}}, \bibinfo
  {author} {\bibfnamefont {J.-H.}\ \bibnamefont {Chu}}, \bibinfo {author}
  {\bibfnamefont {J.~G.}\ \bibnamefont {Analytis}}, \bibinfo {author}
  {\bibfnamefont {A.~P.}\ \bibnamefont {Sorini}}, \bibinfo {author}
  {\bibfnamefont {A.~F.}\ \bibnamefont {Kemper}}, \bibinfo {author}
  {\bibfnamefont {B.}~\bibnamefont {Moritz}}, \bibinfo {author} {\bibfnamefont
  {S.-K.}\ \bibnamefont {Mo}}, \bibinfo {author} {\bibfnamefont {R.~G.}\
  \bibnamefont {Moore}}, \bibinfo {author} {\bibfnamefont {M.}~\bibnamefont
  {Hashimoto}}, \bibinfo {author} {\bibfnamefont {W.-S.}\ \bibnamefont {Lee}},
  \bibinfo {author} {\bibfnamefont {Z.}~\bibnamefont {Hussain}}, \bibinfo
  {author} {\bibfnamefont {T.~P.}\ \bibnamefont {Devereaux}}, \bibinfo {author}
  {\bibfnamefont {I.~R.}\ \bibnamefont {Fisher}}, \ and\ \bibinfo {author}
  {\bibfnamefont {Z.-X.}\ \bibnamefont {Shen}},\ }\href {\doibase
  10.1073/pnas.1015572108} {\bibfield  {journal} {\bibinfo  {journal} {Proc.
  Natl. Acad. Sci. U.S.A.}\ }\textbf {\bibinfo {volume} {108}},\ \bibinfo
  {pages} {6878} (\bibinfo {year} {2011})}\BibitemShut {NoStop}%
\bibitem [{\citenamefont {Wang}\ \emph {et~al.}(2016)\citenamefont {Wang},
  \citenamefont {Feng}, \citenamefont {Cheng}, \citenamefont {Wu},
  \citenamefont {Luo},\ and\ \citenamefont {Rosenbaum}}]{Wang2016}%
  \BibitemOpen
  \bibfield  {author} {\bibinfo {author} {\bibfnamefont {Y.}~\bibnamefont
  {Wang}}, \bibinfo {author} {\bibfnamefont {Y.}~\bibnamefont {Feng}}, \bibinfo
  {author} {\bibfnamefont {J.-G.}\ \bibnamefont {Cheng}}, \bibinfo {author}
  {\bibfnamefont {W.}~\bibnamefont {Wu}}, \bibinfo {author} {\bibfnamefont
  {J.}~\bibnamefont {Luo}}, \ and\ \bibinfo {author} {\bibfnamefont
  {T.}~\bibnamefont {Rosenbaum}},\ }\href@noop {} {\bibfield  {journal}
  {\bibinfo  {journal} {Nat. Commun.}\ }\textbf {\bibinfo {volume} {7}},\
  \bibinfo {pages} {13037} (\bibinfo {year} {2016})}\BibitemShut {NoStop}%
\bibitem [{\citenamefont {Khasanov}\ \emph {et~al.}(2016)\citenamefont
  {Khasanov}, \citenamefont {Amato}, \citenamefont {Bonf\`a}, \citenamefont
  {Guguchia}, \citenamefont {Luetkens}, \citenamefont {Morenzoni},
  \citenamefont {De~Renzi},\ and\ \citenamefont
  {Zhigadlo}}]{PhysRevB.93.180509}%
  \BibitemOpen
  \bibfield  {author} {\bibinfo {author} {\bibfnamefont {R.}~\bibnamefont
  {Khasanov}}, \bibinfo {author} {\bibfnamefont {A.}~\bibnamefont {Amato}},
  \bibinfo {author} {\bibfnamefont {P.}~\bibnamefont {Bonf\`a}}, \bibinfo
  {author} {\bibfnamefont {Z.}~\bibnamefont {Guguchia}}, \bibinfo {author}
  {\bibfnamefont {H.}~\bibnamefont {Luetkens}}, \bibinfo {author}
  {\bibfnamefont {E.}~\bibnamefont {Morenzoni}}, \bibinfo {author}
  {\bibfnamefont {R.}~\bibnamefont {De~Renzi}}, \ and\ \bibinfo {author}
  {\bibfnamefont {N.~D.}\ \bibnamefont {Zhigadlo}},\ }\href
  {https://link.aps.org/doi/10.1103/PhysRevB.93.180509} {\bibfield  {journal}
  {\bibinfo  {journal} {Phys. Rev. B}\ }\textbf {\bibinfo {volume} {93}},\
  \bibinfo {pages} {180509} (\bibinfo {year} {2016})}\BibitemShut {NoStop}%
\bibitem [{\citenamefont {Matsuda}\ \emph {et~al.}(2016)\citenamefont
  {Matsuda}, \citenamefont {Ye}, \citenamefont {Dissanayake}, \citenamefont
  {Cheng}, \citenamefont {Chi}, \citenamefont {Ma}, \citenamefont {Zhou},
  \citenamefont {Yan}, \citenamefont {Kasamatsu}, \citenamefont {Sugino},
  \citenamefont {Kato}, \citenamefont {Matsubayashi}, \citenamefont {Okada},\
  and\ \citenamefont {Uwatoko}}]{PhysRevB.93.100405}%
  \BibitemOpen
  \bibfield  {author} {\bibinfo {author} {\bibfnamefont {M.}~\bibnamefont
  {Matsuda}}, \bibinfo {author} {\bibfnamefont {F.}~\bibnamefont {Ye}},
  \bibinfo {author} {\bibfnamefont {S.~E.}\ \bibnamefont {Dissanayake}},
  \bibinfo {author} {\bibfnamefont {J.-G.}\ \bibnamefont {Cheng}}, \bibinfo
  {author} {\bibfnamefont {S.}~\bibnamefont {Chi}}, \bibinfo {author}
  {\bibfnamefont {J.}~\bibnamefont {Ma}}, \bibinfo {author} {\bibfnamefont
  {H.~D.}\ \bibnamefont {Zhou}}, \bibinfo {author} {\bibfnamefont {J.-Q.}\
  \bibnamefont {Yan}}, \bibinfo {author} {\bibfnamefont {S.}~\bibnamefont
  {Kasamatsu}}, \bibinfo {author} {\bibfnamefont {O.}~\bibnamefont {Sugino}},
  \bibinfo {author} {\bibfnamefont {T.}~\bibnamefont {Kato}}, \bibinfo {author}
  {\bibfnamefont {K.}~\bibnamefont {Matsubayashi}}, \bibinfo {author}
  {\bibfnamefont {T.}~\bibnamefont {Okada}}, \ and\ \bibinfo {author}
  {\bibfnamefont {Y.}~\bibnamefont {Uwatoko}},\ }\href {\doibase
  10.1103/PhysRevB.93.100405} {\bibfield  {journal} {\bibinfo  {journal} {Phys.
  Rev. B}\ }\textbf {\bibinfo {volume} {93}},\ \bibinfo {pages} {100405}
  (\bibinfo {year} {2016})}\BibitemShut {NoStop}%
\bibitem [{\citenamefont {Cheng}\ \emph {et~al.}(2015)\citenamefont {Cheng},
  \citenamefont {Matsubayashi}, \citenamefont {Wu}, \citenamefont {Sun},
  \citenamefont {Lin}, \citenamefont {Luo},\ and\ \citenamefont
  {Uwatoko}}]{cheng2015pressure}%
  \BibitemOpen
  \bibfield  {author} {\bibinfo {author} {\bibfnamefont {J.-G.}\ \bibnamefont
  {Cheng}}, \bibinfo {author} {\bibfnamefont {K.}~\bibnamefont {Matsubayashi}},
  \bibinfo {author} {\bibfnamefont {W.}~\bibnamefont {Wu}}, \bibinfo {author}
  {\bibfnamefont {J.}~\bibnamefont {Sun}}, \bibinfo {author} {\bibfnamefont
  {F.}~\bibnamefont {Lin}}, \bibinfo {author} {\bibfnamefont {J.}~\bibnamefont
  {Luo}}, \ and\ \bibinfo {author} {\bibfnamefont {Y.}~\bibnamefont
  {Uwatoko}},\ }\href {\doibase 10.1103/PhysRevLett.114.117001} {\bibfield
  {journal} {\bibinfo  {journal} {Phys. Rev. Lett.}\ }\textbf {\bibinfo
  {volume} {114}},\ \bibinfo {pages} {117001} (\bibinfo {year}
  {2015})}\BibitemShut {NoStop}%
\bibitem [{\citenamefont {Xu}\ \emph {et~al.}(2017)\citenamefont {Xu},
  \citenamefont {Liu}, \citenamefont {Zheng}, \citenamefont {Chen},
  \citenamefont {Cheng}, \citenamefont {Luo}, \citenamefont {Xie},\ and\
  \citenamefont {Yang}}]{xu2017first}%
  \BibitemOpen
  \bibfield  {author} {\bibinfo {author} {\bibfnamefont {Y.}~\bibnamefont
  {Xu}}, \bibinfo {author} {\bibfnamefont {M.}~\bibnamefont {Liu}}, \bibinfo
  {author} {\bibfnamefont {P.}~\bibnamefont {Zheng}}, \bibinfo {author}
  {\bibfnamefont {X.}~\bibnamefont {Chen}}, \bibinfo {author} {\bibfnamefont
  {J.-g.}\ \bibnamefont {Cheng}}, \bibinfo {author} {\bibfnamefont
  {J.}~\bibnamefont {Luo}}, \bibinfo {author} {\bibfnamefont {W.}~\bibnamefont
  {Xie}}, \ and\ \bibinfo {author} {\bibfnamefont {Y.-f.}\ \bibnamefont
  {Yang}},\ }\href {\doibase 10.1088/1361-648X/aa7023} {\bibfield  {journal}
  {\bibinfo  {journal} {J. Phys.: Condens. Matter}\ }\textbf {\bibinfo {volume}
  {29}},\ \bibinfo {pages} {244001} (\bibinfo {year} {2017})}\BibitemShut
  {NoStop}%
\bibitem [{\citenamefont {De~Groot}\ \emph {et~al.}(1995)\citenamefont
  {De~Groot}, \citenamefont {Pizzini}, \citenamefont {Fontaine}, \citenamefont
  {H{\"a}m{\"a}l{\"a}inen}, \citenamefont {Kao},\ and\ \citenamefont
  {Hastings}}]{de1995local}%
  \BibitemOpen
  \bibfield  {author} {\bibinfo {author} {\bibfnamefont {F.}~\bibnamefont
  {De~Groot}}, \bibinfo {author} {\bibfnamefont {S.}~\bibnamefont {Pizzini}},
  \bibinfo {author} {\bibfnamefont {A.}~\bibnamefont {Fontaine}}, \bibinfo
  {author} {\bibfnamefont {K.}~\bibnamefont {H{\"a}m{\"a}l{\"a}inen}}, \bibinfo
  {author} {\bibfnamefont {C.}~\bibnamefont {Kao}}, \ and\ \bibinfo {author}
  {\bibfnamefont {J.}~\bibnamefont {Hastings}},\ }\href {\doibase
  10.1103/PhysRevB.51.1045} {\bibfield  {journal} {\bibinfo  {journal} {Phys.
  Rev. B}\ }\textbf {\bibinfo {volume} {51}},\ \bibinfo {pages} {1045}
  (\bibinfo {year} {1995})}\BibitemShut {NoStop}%
\bibitem [{\citenamefont {Cuono}\ \emph {et~al.}(2019)\citenamefont {Cuono},
  \citenamefont {Forte}, \citenamefont {Cuoco}, \citenamefont {Islam},
  \citenamefont {Luo}, \citenamefont {Noce},\ and\ \citenamefont
  {Autieri}}]{PhysRevMaterials.3.095004}%
  \BibitemOpen
  \bibfield  {author} {\bibinfo {author} {\bibfnamefont {G.}~\bibnamefont
  {Cuono}}, \bibinfo {author} {\bibfnamefont {F.}~\bibnamefont {Forte}},
  \bibinfo {author} {\bibfnamefont {M.}~\bibnamefont {Cuoco}}, \bibinfo
  {author} {\bibfnamefont {R.}~\bibnamefont {Islam}}, \bibinfo {author}
  {\bibfnamefont {J.}~\bibnamefont {Luo}}, \bibinfo {author} {\bibfnamefont
  {C.}~\bibnamefont {Noce}}, \ and\ \bibinfo {author} {\bibfnamefont
  {C.}~\bibnamefont {Autieri}},\ }\href {\doibase
  10.1103/PhysRevMaterials.3.095004} {\bibfield  {journal} {\bibinfo  {journal}
  {Phys. Rev. Materials}\ }\textbf {\bibinfo {volume} {3}},\ \bibinfo {pages}
  {095004} (\bibinfo {year} {2019})}\BibitemShut {NoStop}%
\bibitem [{\citenamefont {Han}\ \emph {et~al.}(2018)\citenamefont {Han},
  \citenamefont {Wang}, \citenamefont {Wang}, \citenamefont {Li}, \citenamefont
  {Bao}, \citenamefont {Li}, \citenamefont {Botana}, \citenamefont {Xiao},
  \citenamefont {Chow}, \citenamefont {Chung}, \citenamefont {Chen},
  \citenamefont {Wan}, \citenamefont {Kanatzidis}, \citenamefont {Yang},\ and\
  \citenamefont {Mao}}]{Han2018}%
  \BibitemOpen
  \bibfield  {author} {\bibinfo {author} {\bibfnamefont {F.}~\bibnamefont
  {Han}}, \bibinfo {author} {\bibfnamefont {D.}~\bibnamefont {Wang}}, \bibinfo
  {author} {\bibfnamefont {Y.}~\bibnamefont {Wang}}, \bibinfo {author}
  {\bibfnamefont {N.}~\bibnamefont {Li}}, \bibinfo {author} {\bibfnamefont
  {J.-K.}\ \bibnamefont {Bao}}, \bibinfo {author} {\bibfnamefont
  {B.}~\bibnamefont {Li}}, \bibinfo {author} {\bibfnamefont {A.}~\bibnamefont
  {Botana}}, \bibinfo {author} {\bibfnamefont {Y.}~\bibnamefont {Xiao}},
  \bibinfo {author} {\bibfnamefont {P.}~\bibnamefont {Chow}}, \bibinfo {author}
  {\bibfnamefont {D.}~\bibnamefont {Chung}}, \bibinfo {author} {\bibfnamefont
  {J.}~\bibnamefont {Chen}}, \bibinfo {author} {\bibfnamefont {X.}~\bibnamefont
  {Wan}}, \bibinfo {author} {\bibfnamefont {M.}~\bibnamefont {Kanatzidis}},
  \bibinfo {author} {\bibfnamefont {W.}~\bibnamefont {Yang}}, \ and\ \bibinfo
  {author} {\bibfnamefont {H.-K.}\ \bibnamefont {Mao}},\ }\href {\doibase
  10.1088/1367-2630/aaa3c3} {\bibfield  {journal} {\bibinfo  {journal} {New J.
  Phys.}\ }\textbf {\bibinfo {volume} {20}},\ \bibinfo {pages} {023012}
  (\bibinfo {year} {2018})}\BibitemShut {NoStop}%
\bibitem [{\citenamefont {Nozue}\ \emph {et~al.}(2001)\citenamefont {Nozue},
  \citenamefont {Kobayashi}, \citenamefont {Kimura}, \citenamefont {Aoki},\
  and\ \citenamefont {Kamimura}}]{Nozue2001192}%
  \BibitemOpen
  \bibfield  {author} {\bibinfo {author} {\bibfnamefont {T.}~\bibnamefont
  {Nozue}}, \bibinfo {author} {\bibfnamefont {H.}~\bibnamefont {Kobayashi}},
  \bibinfo {author} {\bibfnamefont {N.}~\bibnamefont {Kimura}}, \bibinfo
  {author} {\bibfnamefont {H.}~\bibnamefont {Aoki}}, \ and\ \bibinfo {author}
  {\bibfnamefont {T.}~\bibnamefont {Kamimura}},\ }\href {\doibase
  10.1143/JPSJ.70.192} {\bibfield  {journal} {\bibinfo  {journal} {J. Phys.
  Soc. Jpn.}\ }\textbf {\bibinfo {volume} {70}},\ \bibinfo {pages} {192}
  (\bibinfo {year} {2001})}\BibitemShut {NoStop}%
\bibitem [{\citenamefont {Westerstrandh}\ \emph {et~al.}(1977)\citenamefont
  {Westerstrandh}, \citenamefont {Lundgren}, \citenamefont {G{\"a}fvert},\ and\
  \citenamefont {Carlsson}}]{westerstrandh1977magnetic}%
  \BibitemOpen
  \bibfield  {author} {\bibinfo {author} {\bibfnamefont {B.}~\bibnamefont
  {Westerstrandh}}, \bibinfo {author} {\bibfnamefont {L.}~\bibnamefont
  {Lundgren}}, \bibinfo {author} {\bibfnamefont {U.}~\bibnamefont
  {G{\"a}fvert}}, \ and\ \bibinfo {author} {\bibfnamefont {B.}~\bibnamefont
  {Carlsson}},\ }\href
  {https://iopscience.iop.org/article/10.1088/0031-8949/15/4/009/meta}
  {\bibfield  {journal} {\bibinfo  {journal} {Phys. Scr.}\ }\textbf {\bibinfo
  {volume} {15}},\ \bibinfo {pages} {276} (\bibinfo {year} {1977})}\BibitemShut
  {NoStop}%
\bibitem [{\citenamefont {Kawakatsu}\ \emph {et~al.}(2019)\citenamefont
  {Kawakatsu}, \citenamefont {Kakihana}, \citenamefont {Nakashima},
  \citenamefont {Amako}, \citenamefont {Nakamura}, \citenamefont {Aoki},
  \citenamefont {Takeuchi}, \citenamefont {Harima}, \citenamefont {Hedo},
  \citenamefont {Nakama} \emph {et~al.}}]{kawakatsu2019haas}%
  \BibitemOpen
  \bibfield  {author} {\bibinfo {author} {\bibfnamefont {S.}~\bibnamefont
  {Kawakatsu}}, \bibinfo {author} {\bibfnamefont {M.}~\bibnamefont {Kakihana}},
  \bibinfo {author} {\bibfnamefont {M.}~\bibnamefont {Nakashima}}, \bibinfo
  {author} {\bibfnamefont {Y.}~\bibnamefont {Amako}}, \bibinfo {author}
  {\bibfnamefont {A.}~\bibnamefont {Nakamura}}, \bibinfo {author}
  {\bibfnamefont {D.}~\bibnamefont {Aoki}}, \bibinfo {author} {\bibfnamefont
  {T.}~\bibnamefont {Takeuchi}}, \bibinfo {author} {\bibfnamefont
  {H.}~\bibnamefont {Harima}}, \bibinfo {author} {\bibfnamefont
  {M.}~\bibnamefont {Hedo}}, \bibinfo {author} {\bibfnamefont {T.}~\bibnamefont
  {Nakama}},  \emph {et~al.},\ }\href {\doibase 10.7566/JPSJ.88.044705}
  {\bibfield  {journal} {\bibinfo  {journal} {J. Phys. Soc. Jpn.}\ }\textbf
  {\bibinfo {volume} {88}},\ \bibinfo {pages} {044705} (\bibinfo {year}
  {2019})}\BibitemShut {NoStop}%
\bibitem [{\citenamefont {Yin}\ \emph {et~al.}(2019)\citenamefont {Yin},
  \citenamefont {Zhai}, \citenamefont {Zhang},\ and\ \citenamefont
  {Zhai}}]{yin2019electrical}%
  \BibitemOpen
  \bibfield  {author} {\bibinfo {author} {\bibfnamefont {Y.}~\bibnamefont
  {Yin}}, \bibinfo {author} {\bibfnamefont {K.}~\bibnamefont {Zhai}}, \bibinfo
  {author} {\bibfnamefont {B.}~\bibnamefont {Zhang}}, \ and\ \bibinfo {author}
  {\bibfnamefont {S.}~\bibnamefont {Zhai}},\ }\href@noop {} {\bibfield
  {journal} {\bibinfo  {journal} {J. Geophys. Res. B}\ } (\bibinfo {year}
  {2019})}\BibitemShut {NoStop}%
\bibitem [{\citenamefont {Shiomi}\ \emph {et~al.}(2012)\citenamefont {Shiomi},
  \citenamefont {Iguchi},\ and\ \citenamefont {Tokura}}]{PhysRevB.86.180404}%
  \BibitemOpen
  \bibfield  {author} {\bibinfo {author} {\bibfnamefont {Y.}~\bibnamefont
  {Shiomi}}, \bibinfo {author} {\bibfnamefont {S.}~\bibnamefont {Iguchi}}, \
  and\ \bibinfo {author} {\bibfnamefont {Y.}~\bibnamefont {Tokura}},\ }\href
  {\doibase 10.1103/PhysRevB.86.180404} {\bibfield  {journal} {\bibinfo
  {journal} {Phys. Rev. B}\ }\textbf {\bibinfo {volume} {86}},\ \bibinfo
  {pages} {180404} (\bibinfo {year} {2012})}\BibitemShut {NoStop}%
\bibitem [{\citenamefont {Yano}\ \emph {et~al.}(2018)\citenamefont {Yano},
  \citenamefont {Lan{\c{c}}on}, \citenamefont {R{\o}nnow}, \citenamefont
  {Hansen}, \citenamefont {Ressouche}, \citenamefont {Qureshi}, \citenamefont
  {Ouladdiaf},\ and\ \citenamefont {Gardner}}]{yano2018suppression}%
  \BibitemOpen
  \bibfield  {author} {\bibinfo {author} {\bibfnamefont {S.-i.}\ \bibnamefont
  {Yano}}, \bibinfo {author} {\bibfnamefont {D.}~\bibnamefont {Lan{\c{c}}on}},
  \bibinfo {author} {\bibfnamefont {H.~M.}\ \bibnamefont {R{\o}nnow}}, \bibinfo
  {author} {\bibfnamefont {T.~C.}\ \bibnamefont {Hansen}}, \bibinfo {author}
  {\bibfnamefont {E.}~\bibnamefont {Ressouche}}, \bibinfo {author}
  {\bibfnamefont {N.}~\bibnamefont {Qureshi}}, \bibinfo {author} {\bibfnamefont
  {B.}~\bibnamefont {Ouladdiaf}}, \ and\ \bibinfo {author} {\bibfnamefont
  {J.~S.}\ \bibnamefont {Gardner}},\ }\href {\doibase 10.7566/JPSJ.87.023703}
  {\bibfield  {journal} {\bibinfo  {journal} {J. Phys. Soc. Jpn.}\ }\textbf
  {\bibinfo {volume} {87}},\ \bibinfo {pages} {023703} (\bibinfo {year}
  {2018})}\BibitemShut {NoStop}%
\bibitem [{\citenamefont {Zhigadlo}\ \emph {et~al.}(2017)\citenamefont
  {Zhigadlo}, \citenamefont {Barbero},\ and\ \citenamefont
  {Shiroka}}]{zhigadlo2017growth}%
  \BibitemOpen
  \bibfield  {author} {\bibinfo {author} {\bibfnamefont {N.~D.}\ \bibnamefont
  {Zhigadlo}}, \bibinfo {author} {\bibfnamefont {N.}~\bibnamefont {Barbero}}, \
  and\ \bibinfo {author} {\bibfnamefont {T.}~\bibnamefont {Shiroka}},\ }\href
  {\doibase 10.1016/j.jallcom.2017.07.247} {\bibfield  {journal} {\bibinfo
  {journal} {J. Alloys Compd.}\ }\textbf {\bibinfo {volume} {725}},\ \bibinfo
  {pages} {1027} (\bibinfo {year} {2017})}\BibitemShut {NoStop}%
\bibitem [{\citenamefont {T{\'o}th}\ \emph {et~al.}(2016)\citenamefont
  {T{\'o}th}, \citenamefont {Wehinger}, \citenamefont {Rolfs}, \citenamefont
  {Birol}, \citenamefont {Stuhr}, \citenamefont {Takatsu}, \citenamefont
  {Kimura}, \citenamefont {Kimura}, \citenamefont {R{\o}nnow},\ and\
  \citenamefont {R{\"u}egg}}]{toth2016electromagnon}%
  \BibitemOpen
  \bibfield  {author} {\bibinfo {author} {\bibfnamefont {S.}~\bibnamefont
  {T{\'o}th}}, \bibinfo {author} {\bibfnamefont {B.}~\bibnamefont {Wehinger}},
  \bibinfo {author} {\bibfnamefont {K.}~\bibnamefont {Rolfs}}, \bibinfo
  {author} {\bibfnamefont {T.}~\bibnamefont {Birol}}, \bibinfo {author}
  {\bibfnamefont {U.}~\bibnamefont {Stuhr}}, \bibinfo {author} {\bibfnamefont
  {H.}~\bibnamefont {Takatsu}}, \bibinfo {author} {\bibfnamefont
  {K.}~\bibnamefont {Kimura}}, \bibinfo {author} {\bibfnamefont
  {T.}~\bibnamefont {Kimura}}, \bibinfo {author} {\bibfnamefont {H.~M.}\
  \bibnamefont {R{\o}nnow}}, \ and\ \bibinfo {author} {\bibfnamefont
  {C.}~\bibnamefont {R{\"u}egg}},\ }\href {\doibase 10.1038/ncomms13547}
  {\bibfield  {journal} {\bibinfo  {journal} {Nat. Commun.}\ }\textbf {\bibinfo
  {volume} {7}},\ \bibinfo {pages} {13547} (\bibinfo {year}
  {2016})}\BibitemShut {NoStop}%
\bibitem [{\citenamefont {Oh}\ \emph {et~al.}(2016)\citenamefont {Oh},
  \citenamefont {Le}, \citenamefont {Nahm}, \citenamefont {Sim}, \citenamefont
  {Jeong}, \citenamefont {Perring}, \citenamefont {Woo}, \citenamefont
  {Nakajima}, \citenamefont {Ohira-Kawamura}, \citenamefont {Yamani} \emph
  {et~al.}}]{oh2016spontaneous}%
  \BibitemOpen
  \bibfield  {author} {\bibinfo {author} {\bibfnamefont {J.}~\bibnamefont
  {Oh}}, \bibinfo {author} {\bibfnamefont {M.~D.}\ \bibnamefont {Le}}, \bibinfo
  {author} {\bibfnamefont {H.-H.}\ \bibnamefont {Nahm}}, \bibinfo {author}
  {\bibfnamefont {H.}~\bibnamefont {Sim}}, \bibinfo {author} {\bibfnamefont
  {J.}~\bibnamefont {Jeong}}, \bibinfo {author} {\bibfnamefont
  {T.}~\bibnamefont {Perring}}, \bibinfo {author} {\bibfnamefont
  {H.}~\bibnamefont {Woo}}, \bibinfo {author} {\bibfnamefont {K.}~\bibnamefont
  {Nakajima}}, \bibinfo {author} {\bibfnamefont {S.}~\bibnamefont
  {Ohira-Kawamura}}, \bibinfo {author} {\bibfnamefont {Z.}~\bibnamefont
  {Yamani}},  \emph {et~al.},\ }\href {\doibase 10.1038/ncomms13146} {\bibfield
   {journal} {\bibinfo  {journal} {Nat. Commun.}\ }\textbf {\bibinfo {volume}
  {7}},\ \bibinfo {pages} {13146} (\bibinfo {year} {2016})}\BibitemShut
  {NoStop}%
\bibitem [{\citenamefont {{\L}a{\.z}ewski}\ \emph {et~al.}(2010)\citenamefont
  {{\L}a{\.z}ewski}, \citenamefont {Piekarz}, \citenamefont {Tobo{\l}a},
  \citenamefont {Wiendlocha}, \citenamefont {Jochym}, \citenamefont {Sternik},\
  and\ \citenamefont {Parlinski}}]{lazewski2010phonon}%
  \BibitemOpen
  \bibfield  {author} {\bibinfo {author} {\bibfnamefont {J.}~\bibnamefont
  {{\L}a{\.z}ewski}}, \bibinfo {author} {\bibfnamefont {P.}~\bibnamefont
  {Piekarz}}, \bibinfo {author} {\bibfnamefont {J.}~\bibnamefont {Tobo{\l}a}},
  \bibinfo {author} {\bibfnamefont {B.}~\bibnamefont {Wiendlocha}}, \bibinfo
  {author} {\bibfnamefont {P.}~\bibnamefont {Jochym}}, \bibinfo {author}
  {\bibfnamefont {M.}~\bibnamefont {Sternik}}, \ and\ \bibinfo {author}
  {\bibfnamefont {K.}~\bibnamefont {Parlinski}},\ }\href {\doibase
  10.1103/PhysRevLett.104.147205} {\bibfield  {journal} {\bibinfo  {journal}
  {Phys. Rev. Lett.}\ }\textbf {\bibinfo {volume} {104}},\ \bibinfo {pages}
  {147205} (\bibinfo {year} {2010})}\BibitemShut {NoStop}%
\bibitem [{\citenamefont {\L{}a\ifmmode~\dot{z}\else \.{z}\fi{}ewski}\ \emph
  {et~al.}(2011)\citenamefont {\L{}a\ifmmode~\dot{z}\else \.{z}\fi{}ewski},
  \citenamefont {Piekarz},\ and\ \citenamefont
  {Parlinski}}]{PhysRevB.83.054108}%
  \BibitemOpen
  \bibfield  {author} {\bibinfo {author} {\bibfnamefont {J.}~\bibnamefont
  {\L{}a\ifmmode~\dot{z}\else \.{z}\fi{}ewski}}, \bibinfo {author}
  {\bibfnamefont {P.}~\bibnamefont {Piekarz}}, \ and\ \bibinfo {author}
  {\bibfnamefont {K.}~\bibnamefont {Parlinski}},\ }\href {\doibase
  10.1103/PhysRevB.83.054108} {\bibfield  {journal} {\bibinfo  {journal} {Phys.
  Rev. B}\ }\textbf {\bibinfo {volume} {83}},\ \bibinfo {pages} {054108}
  (\bibinfo {year} {2011})}\BibitemShut {NoStop}%
\bibitem [{\citenamefont {Rungger}\ and\ \citenamefont
  {Sanvito}(2006)}]{PhysRevB.74.024429}%
  \BibitemOpen
  \bibfield  {author} {\bibinfo {author} {\bibfnamefont {I.}~\bibnamefont
  {Rungger}}\ and\ \bibinfo {author} {\bibfnamefont {S.}~\bibnamefont
  {Sanvito}},\ }\href {\doibase 10.1103/PhysRevB.74.024429} {\bibfield
  {journal} {\bibinfo  {journal} {Phys. Rev. B}\ }\textbf {\bibinfo {volume}
  {74}},\ \bibinfo {pages} {024429} (\bibinfo {year} {2006})}\BibitemShut
  {NoStop}%
\bibitem [{\citenamefont {Boller}\ and\ \citenamefont
  {Kallel}(1971)}]{BOLLER19711699}%
  \BibitemOpen
  \bibfield  {author} {\bibinfo {author} {\bibfnamefont {H.}~\bibnamefont
  {Boller}}\ and\ \bibinfo {author} {\bibfnamefont {A.}~\bibnamefont
  {Kallel}},\ }\href {\doibase https://doi.org/10.1016/0038-1098(71)90344-9}
  {\bibfield  {journal} {\bibinfo  {journal} {Solid State Commun.}\ }\textbf
  {\bibinfo {volume} {9}},\ \bibinfo {pages} {1699 } (\bibinfo {year}
  {1971})}\BibitemShut {NoStop}%
\bibitem [{\citenamefont {Kazama}\ and\ \citenamefont
  {Watanabe}(1971)}]{doi:10.1143/JPSJ.30.1319}%
  \BibitemOpen
  \bibfield  {author} {\bibinfo {author} {\bibfnamefont {N.}~\bibnamefont
  {Kazama}}\ and\ \bibinfo {author} {\bibfnamefont {H.}~\bibnamefont
  {Watanabe}},\ }\href@noop {} {\bibfield  {journal} {\bibinfo  {journal} {J.
  Phys. Soc. Jpn.}\ }\textbf {\bibinfo {volume} {30}},\ \bibinfo {pages} {1319}
  (\bibinfo {year} {1971})},\ \Eprint
  {http://arxiv.org/abs/https://doi.org/10.1143/JPSJ.30.1319}
  {https://doi.org/10.1143/JPSJ.30.1319} \BibitemShut {NoStop}%
\bibitem [{\citenamefont {Wu}\ \emph {et~al.}(2014)\citenamefont {Wu},
  \citenamefont {Cheng}, \citenamefont {Matsubayashi}, \citenamefont {Kong},
  \citenamefont {Lin}, \citenamefont {Jin}, \citenamefont {Wang}, \citenamefont
  {Uwatoko},\ and\ \citenamefont {Luo}}]{Wu2014}%
  \BibitemOpen
  \bibfield  {author} {\bibinfo {author} {\bibfnamefont {W.}~\bibnamefont
  {Wu}}, \bibinfo {author} {\bibfnamefont {J.}~\bibnamefont {Cheng}}, \bibinfo
  {author} {\bibfnamefont {K.}~\bibnamefont {Matsubayashi}}, \bibinfo {author}
  {\bibfnamefont {P.}~\bibnamefont {Kong}}, \bibinfo {author} {\bibfnamefont
  {F.}~\bibnamefont {Lin}}, \bibinfo {author} {\bibfnamefont {C.}~\bibnamefont
  {Jin}}, \bibinfo {author} {\bibfnamefont {N.}~\bibnamefont {Wang}}, \bibinfo
  {author} {\bibfnamefont {Y.}~\bibnamefont {Uwatoko}}, \ and\ \bibinfo
  {author} {\bibfnamefont {J.}~\bibnamefont {Luo}},\ }\href
  {https://www.scopus.com/inward/record.uri?eid=2-s2.0-84922252072&doi=10.1038%2fncomms6508&partnerID=40&md5=e7f8c7fc0671f1c1ce5cd28af61aa0d1}
  {\bibfield  {journal} {\bibinfo  {journal} {Nat. Commun.}\ }\textbf {\bibinfo
  {volume} {5}},\ \bibinfo {pages} {5508} (\bibinfo {year} {2014})}\BibitemShut
  {NoStop}%
\bibitem [{\citenamefont {Guo}\ \emph {et~al.}(2018)\citenamefont {Guo},
  \citenamefont {Smidman}, \citenamefont {Shen}, \citenamefont {Wu},
  \citenamefont {Lin}, \citenamefont {Han}, \citenamefont {Chen}, \citenamefont
  {Wu}, \citenamefont {Wang}, \citenamefont {Jiang}, \citenamefont {Lu},
  \citenamefont {Hu}, \citenamefont {Luo},\ and\ \citenamefont
  {Yuan}}]{PhysRevB.98.024520}%
  \BibitemOpen
  \bibfield  {author} {\bibinfo {author} {\bibfnamefont {C.~Y.}\ \bibnamefont
  {Guo}}, \bibinfo {author} {\bibfnamefont {M.}~\bibnamefont {Smidman}},
  \bibinfo {author} {\bibfnamefont {B.}~\bibnamefont {Shen}}, \bibinfo {author}
  {\bibfnamefont {W.}~\bibnamefont {Wu}}, \bibinfo {author} {\bibfnamefont
  {F.~K.}\ \bibnamefont {Lin}}, \bibinfo {author} {\bibfnamefont {X.~L.}\
  \bibnamefont {Han}}, \bibinfo {author} {\bibfnamefont {Y.}~\bibnamefont
  {Chen}}, \bibinfo {author} {\bibfnamefont {F.}~\bibnamefont {Wu}}, \bibinfo
  {author} {\bibfnamefont {Y.~F.}\ \bibnamefont {Wang}}, \bibinfo {author}
  {\bibfnamefont {W.~B.}\ \bibnamefont {Jiang}}, \bibinfo {author}
  {\bibfnamefont {X.}~\bibnamefont {Lu}}, \bibinfo {author} {\bibfnamefont
  {J.~P.}\ \bibnamefont {Hu}}, \bibinfo {author} {\bibfnamefont {J.~L.}\
  \bibnamefont {Luo}}, \ and\ \bibinfo {author} {\bibfnamefont {H.~Q.}\
  \bibnamefont {Yuan}},\ }\href {\doibase 10.1103/PhysRevB.98.024520}
  {\bibfield  {journal} {\bibinfo  {journal} {Phys. Rev. B}\ }\textbf {\bibinfo
  {volume} {98}},\ \bibinfo {pages} {024520} (\bibinfo {year}
  {2018})}\BibitemShut {NoStop}%
\bibitem [{\citenamefont {Kotegawa}\ \emph {et~al.}(2015)\citenamefont
  {Kotegawa}, \citenamefont {Nakahara}, \citenamefont {Akamatsu}, \citenamefont
  {Tou}, \citenamefont {Sugawara},\ and\ \citenamefont
  {Harima}}]{PhysRevLett.114.117002}%
  \BibitemOpen
  \bibfield  {author} {\bibinfo {author} {\bibfnamefont {H.}~\bibnamefont
  {Kotegawa}}, \bibinfo {author} {\bibfnamefont {S.}~\bibnamefont {Nakahara}},
  \bibinfo {author} {\bibfnamefont {R.}~\bibnamefont {Akamatsu}}, \bibinfo
  {author} {\bibfnamefont {H.}~\bibnamefont {Tou}}, \bibinfo {author}
  {\bibfnamefont {H.}~\bibnamefont {Sugawara}}, \ and\ \bibinfo {author}
  {\bibfnamefont {H.}~\bibnamefont {Harima}},\ }\href {\doibase
  10.1103/PhysRevLett.114.117002} {\bibfield  {journal} {\bibinfo  {journal}
  {Phys. Rev. Lett.}\ }\textbf {\bibinfo {volume} {114}},\ \bibinfo {pages}
  {117002} (\bibinfo {year} {2015})}\BibitemShut {NoStop}%
\bibitem [{\citenamefont {Matsuda}\ \emph {et~al.}(2018)\citenamefont
  {Matsuda}, \citenamefont {Lin}, \citenamefont {Yu}, \citenamefont {Cheng},
  \citenamefont {Wu}, \citenamefont {Sun}, \citenamefont {Zhang}, \citenamefont
  {Sun}, \citenamefont {Matsubayashi}, \citenamefont {Miyake}, \citenamefont
  {Kato}, \citenamefont {Yan}, \citenamefont {Stone}, \citenamefont {Si},
  \citenamefont {Luo},\ and\ \citenamefont {Uwatoko}}]{PhysRevX.8.031017}%
  \BibitemOpen
  \bibfield  {author} {\bibinfo {author} {\bibfnamefont {M.}~\bibnamefont
  {Matsuda}}, \bibinfo {author} {\bibfnamefont {F.~K.}\ \bibnamefont {Lin}},
  \bibinfo {author} {\bibfnamefont {R.}~\bibnamefont {Yu}}, \bibinfo {author}
  {\bibfnamefont {J.-G.}\ \bibnamefont {Cheng}}, \bibinfo {author}
  {\bibfnamefont {W.}~\bibnamefont {Wu}}, \bibinfo {author} {\bibfnamefont
  {J.~P.}\ \bibnamefont {Sun}}, \bibinfo {author} {\bibfnamefont {J.~H.}\
  \bibnamefont {Zhang}}, \bibinfo {author} {\bibfnamefont {P.~J.}\ \bibnamefont
  {Sun}}, \bibinfo {author} {\bibfnamefont {K.}~\bibnamefont {Matsubayashi}},
  \bibinfo {author} {\bibfnamefont {T.}~\bibnamefont {Miyake}}, \bibinfo
  {author} {\bibfnamefont {T.}~\bibnamefont {Kato}}, \bibinfo {author}
  {\bibfnamefont {J.-Q.}\ \bibnamefont {Yan}}, \bibinfo {author} {\bibfnamefont
  {M.~B.}\ \bibnamefont {Stone}}, \bibinfo {author} {\bibfnamefont
  {Q.}~\bibnamefont {Si}}, \bibinfo {author} {\bibfnamefont {J.~L.}\
  \bibnamefont {Luo}}, \ and\ \bibinfo {author} {\bibfnamefont
  {Y.}~\bibnamefont {Uwatoko}},\ }\href {\doibase 10.1103/PhysRevX.8.031017}
  {\bibfield  {journal} {\bibinfo  {journal} {Phys. Rev. X}\ }\textbf {\bibinfo
  {volume} {8}},\ \bibinfo {pages} {031017} (\bibinfo {year}
  {2018})}\BibitemShut {NoStop}%
\bibitem [{\citenamefont {Park}\ \emph {et~al.}(2019)\citenamefont {Park},
  \citenamefont {Shin}, \citenamefont {Kim}, \citenamefont {Kim}, \citenamefont
  {Park}, \citenamefont {Thompson},\ and\ \citenamefont {Park}}]{Park2019}%
  \BibitemOpen
  \bibfield  {author} {\bibinfo {author} {\bibfnamefont {S.}~\bibnamefont
  {Park}}, \bibinfo {author} {\bibfnamefont {S.}~\bibnamefont {Shin}}, \bibinfo
  {author} {\bibfnamefont {S.-I.}\ \bibnamefont {Kim}}, \bibinfo {author}
  {\bibfnamefont {S.}~\bibnamefont {Kim}}, \bibinfo {author} {\bibfnamefont
  {C.-K.}\ \bibnamefont {Park}}, \bibinfo {author} {\bibfnamefont
  {J.}~\bibnamefont {Thompson}}, \ and\ \bibinfo {author} {\bibfnamefont
  {T.}~\bibnamefont {Park}},\ }\href
  {https://www.scopus.com/inward/record.uri?eid=2-s2.0-85071927107&doi=10.1038%2fs41535-019-0188-6&partnerID=40&md5=70e6c53de8af6d4d6de006b17625d00b}
  {\bibfield  {journal} {\bibinfo  {journal} {npj Quantum Mater.}\ }\textbf
  {\bibinfo {volume} {4}} (\bibinfo {year} {2019})}\BibitemShut {NoStop}%
\bibitem [{\citenamefont {Shen}\ \emph {et~al.}(2016)\citenamefont {Shen},
  \citenamefont {Wang}, \citenamefont {Hao}, \citenamefont {Pan}, \citenamefont
  {Feng}, \citenamefont {Huang}, \citenamefont {Harriger}, \citenamefont
  {Leao}, \citenamefont {Zhao}, \citenamefont {Chisnell}, \citenamefont {Lynn},
  \citenamefont {Cao}, \citenamefont {Hu},\ and\ \citenamefont
  {Zhao}}]{PhysRevB.93.060503}%
  \BibitemOpen
  \bibfield  {author} {\bibinfo {author} {\bibfnamefont {Y.}~\bibnamefont
  {Shen}}, \bibinfo {author} {\bibfnamefont {Q.}~\bibnamefont {Wang}}, \bibinfo
  {author} {\bibfnamefont {Y.}~\bibnamefont {Hao}}, \bibinfo {author}
  {\bibfnamefont {B.}~\bibnamefont {Pan}}, \bibinfo {author} {\bibfnamefont
  {Y.}~\bibnamefont {Feng}}, \bibinfo {author} {\bibfnamefont {Q.}~\bibnamefont
  {Huang}}, \bibinfo {author} {\bibfnamefont {L.~W.}\ \bibnamefont {Harriger}},
  \bibinfo {author} {\bibfnamefont {J.~B.}\ \bibnamefont {Leao}}, \bibinfo
  {author} {\bibfnamefont {Y.}~\bibnamefont {Zhao}}, \bibinfo {author}
  {\bibfnamefont {R.~M.}\ \bibnamefont {Chisnell}}, \bibinfo {author}
  {\bibfnamefont {J.~W.}\ \bibnamefont {Lynn}}, \bibinfo {author}
  {\bibfnamefont {H.}~\bibnamefont {Cao}}, \bibinfo {author} {\bibfnamefont
  {J.}~\bibnamefont {Hu}}, \ and\ \bibinfo {author} {\bibfnamefont
  {J.}~\bibnamefont {Zhao}},\ }\href {\doibase 10.1103/PhysRevB.93.060503}
  {\bibfield  {journal} {\bibinfo  {journal} {Phys. Rev. B}\ }\textbf {\bibinfo
  {volume} {93}},\ \bibinfo {pages} {060503} (\bibinfo {year}
  {2016})}\BibitemShut {NoStop}%
\bibitem [{\citenamefont {Chernyavskii}\ \emph {et~al.}(2020)\citenamefont
  {Chernyavskii}, \citenamefont {Nikitin}, \citenamefont {Onykiienko},
  \citenamefont {Inosov}, \citenamefont {Stahl}, \citenamefont {Geck},
  \citenamefont {Hong}, \citenamefont {Hess}, \citenamefont {Gass},
  \citenamefont {Wolter}, \citenamefont {Wolf}, \citenamefont {Lubk},
  \citenamefont {Efremov}, \citenamefont {Yokaichiya}, \citenamefont
  {Aswartham}, \citenamefont {B\"uchner},\ and\ \citenamefont
  {Morozov}}]{Chernyavskii}%
  \BibitemOpen
  \bibfield  {author} {\bibinfo {author} {\bibfnamefont {I.~O.}\ \bibnamefont
  {Chernyavskii}}, \bibinfo {author} {\bibfnamefont {S.~E.}\ \bibnamefont
  {Nikitin}}, \bibinfo {author} {\bibfnamefont {Y.~A.}\ \bibnamefont
  {Onykiienko}}, \bibinfo {author} {\bibfnamefont {D.~S.}\ \bibnamefont
  {Inosov}}, \bibinfo {author} {\bibfnamefont {Q.}~\bibnamefont {Stahl}},
  \bibinfo {author} {\bibfnamefont {J.}~\bibnamefont {Geck}}, \bibinfo {author}
  {\bibfnamefont {X.~C.}\ \bibnamefont {Hong}}, \bibinfo {author}
  {\bibfnamefont {C.}~\bibnamefont {Hess}}, \bibinfo {author} {\bibfnamefont
  {S.}~\bibnamefont {Gass}}, \bibinfo {author} {\bibfnamefont {A.~U.~B.}\
  \bibnamefont {Wolter}}, \bibinfo {author} {\bibfnamefont {D.}~\bibnamefont
  {Wolf}}, \bibinfo {author} {\bibfnamefont {A.}~\bibnamefont {Lubk}}, \bibinfo
  {author} {\bibfnamefont {D.~V.}\ \bibnamefont {Efremov}}, \bibinfo {author}
  {\bibfnamefont {F.}~\bibnamefont {Yokaichiya}}, \bibinfo {author}
  {\bibfnamefont {S.}~\bibnamefont {Aswartham}}, \bibinfo {author}
  {\bibfnamefont {B.}~\bibnamefont {B\"uchner}}, \ and\ \bibinfo {author}
  {\bibfnamefont {I.~V.}\ \bibnamefont {Morozov}},\ }\href
  {https://link.aps.org/doi/10.1103/PhysRevMaterials.4.083403} {\bibfield
  {journal} {\bibinfo  {journal} {Phys. Rev. Materials}\ }\textbf {\bibinfo
  {volume} {4}},\ \bibinfo {pages} {083403} (\bibinfo {year}
  {2020})}\BibitemShut {NoStop}%
\bibitem [{\citenamefont {Bewley}\ \emph {et~al.}(2006)\citenamefont {Bewley},
  \citenamefont {Eccleston}, \citenamefont {McEwen}, \citenamefont {Hayden},
  \citenamefont {Dove}, \citenamefont {Bennington}, \citenamefont {Treadgold},\
  and\ \citenamefont {Coleman}}]{BEWLEY20061029}%
  \BibitemOpen
  \bibfield  {author} {\bibinfo {author} {\bibfnamefont {R.}~\bibnamefont
  {Bewley}}, \bibinfo {author} {\bibfnamefont {R.}~\bibnamefont {Eccleston}},
  \bibinfo {author} {\bibfnamefont {K.}~\bibnamefont {McEwen}}, \bibinfo
  {author} {\bibfnamefont {S.}~\bibnamefont {Hayden}}, \bibinfo {author}
  {\bibfnamefont {M.}~\bibnamefont {Dove}}, \bibinfo {author} {\bibfnamefont
  {S.}~\bibnamefont {Bennington}}, \bibinfo {author} {\bibfnamefont
  {J.}~\bibnamefont {Treadgold}}, \ and\ \bibinfo {author} {\bibfnamefont
  {R.}~\bibnamefont {Coleman}},\ }\href {\doibase
  https://doi.org/10.1016/j.physb.2006.05.328} {\bibfield  {journal} {\bibinfo
  {journal} {Physica B}\ }\textbf {\bibinfo {volume} {385-386}},\ \bibinfo
  {pages} {1029 } (\bibinfo {year} {2006})}\BibitemShut {NoStop}%
\bibitem [{\citenamefont {Ewings}\ \emph {et~al.}(2016)\citenamefont {Ewings},
  \citenamefont {Buts}, \citenamefont {Le}, \citenamefont {van Duijn},
  \citenamefont {Bustinduy},\ and\ \citenamefont {Perring}}]{EWINGS2016132}%
  \BibitemOpen
  \bibfield  {author} {\bibinfo {author} {\bibfnamefont {R.}~\bibnamefont
  {Ewings}}, \bibinfo {author} {\bibfnamefont {A.}~\bibnamefont {Buts}},
  \bibinfo {author} {\bibfnamefont {M.}~\bibnamefont {Le}}, \bibinfo {author}
  {\bibfnamefont {J.}~\bibnamefont {van Duijn}}, \bibinfo {author}
  {\bibfnamefont {I.}~\bibnamefont {Bustinduy}}, \ and\ \bibinfo {author}
  {\bibfnamefont {T.}~\bibnamefont {Perring}},\ }\href {\doibase
  https://doi.org/10.1016/j.nima.2016.07.036} {\bibfield  {journal} {\bibinfo
  {journal} {Nucl. Instrum. Methods Phys. Res. A}\ }\textbf {\bibinfo {volume}
  {834}},\ \bibinfo {pages} {132 } (\bibinfo {year} {2016})}\BibitemShut
  {NoStop}%
\bibitem [{\citenamefont
  {Reznik}()}]{https://github.com/dmitryr1234/phonon-explorer}%
  \BibitemOpen
  \bibfield  {author} {\bibinfo {author} {\bibfnamefont {D.}~\bibnamefont
  {Reznik}},\ }\href {https://github.com/dmitryr1234/phonon-explorer} {\bibinfo
   {journal} {https://github.com/dmitryr1234/phonon-explorer}\ }\BibitemShut
  {NoStop}%
\bibitem [{\citenamefont {Kresse}\ and\ \citenamefont
  {Joubert}(1999)}]{PhysRevB.59.1758}%
  \BibitemOpen
\bibfield  {journal} {  }\bibfield  {author} {\bibinfo {author} {\bibfnamefont
  {G.}~\bibnamefont {Kresse}}\ and\ \bibinfo {author} {\bibfnamefont
  {D.}~\bibnamefont {Joubert}},\ }\href {\doibase 10.1103/PhysRevB.59.1758}
  {\bibfield  {journal} {\bibinfo  {journal} {Phys. Rev. B}\ }\textbf {\bibinfo
  {volume} {59}},\ \bibinfo {pages} {1758} (\bibinfo {year}
  {1999})}\BibitemShut {NoStop}%
\bibitem [{\citenamefont {Kresse}\ and\ \citenamefont
  {Furthm{\"u}ller}(1996{\natexlab{a}})}]{PhysRevB.54.11169}%
  \BibitemOpen
  \bibfield  {author} {\bibinfo {author} {\bibfnamefont {G.}~\bibnamefont
  {Kresse}}\ and\ \bibinfo {author} {\bibfnamefont {J.}~\bibnamefont
  {Furthm{\"u}ller}},\ }\href {\doibase 10.1103/PhysRevB.54.11169} {\bibfield
  {journal} {\bibinfo  {journal} {Phys. Rev. B}\ }\textbf {\bibinfo {volume}
  {54}},\ \bibinfo {pages} {11169} (\bibinfo {year}
  {1996}{\natexlab{a}})}\BibitemShut {NoStop}%
\bibitem [{\citenamefont {Kresse}\ and\ \citenamefont
  {Furthm{\"u}ller}(1996{\natexlab{b}})}]{KRESSE199615}%
  \BibitemOpen
  \bibfield  {author} {\bibinfo {author} {\bibfnamefont {G.}~\bibnamefont
  {Kresse}}\ and\ \bibinfo {author} {\bibfnamefont {J.}~\bibnamefont
  {Furthm{\"u}ller}},\ }\href {\doibase
  https://doi.org/10.1016/0927-0256(96)00008-0} {\bibfield  {journal} {\bibinfo
   {journal} {Comput. Mater. Sci.}\ }\textbf {\bibinfo {volume} {6}},\ \bibinfo
  {pages} {15 } (\bibinfo {year} {1996}{\natexlab{b}})}\BibitemShut {NoStop}%
\bibitem [{\citenamefont {Perdew}\ \emph {et~al.}(1996)\citenamefont {Perdew},
  \citenamefont {Burke},\ and\ \citenamefont
  {Ernzerhof}}]{PhysRevLett.77.3865}%
  \BibitemOpen
  \bibfield  {author} {\bibinfo {author} {\bibfnamefont {J.~P.}\ \bibnamefont
  {Perdew}}, \bibinfo {author} {\bibfnamefont {K.}~\bibnamefont {Burke}}, \
  and\ \bibinfo {author} {\bibfnamefont {M.}~\bibnamefont {Ernzerhof}},\ }\href
  {\doibase 10.1103/PhysRevLett.77.3865} {\bibfield  {journal} {\bibinfo
  {journal} {Phys. Rev. Lett.}\ }\textbf {\bibinfo {volume} {77}},\ \bibinfo
  {pages} {3865} (\bibinfo {year} {1996})}\BibitemShut {NoStop}%
\bibitem [{\citenamefont {Monkhorst}\ and\ \citenamefont
  {Pack}(1976)}]{PhysRevB.13.5188}%
  \BibitemOpen
  \bibfield  {author} {\bibinfo {author} {\bibfnamefont {H.~J.}\ \bibnamefont
  {Monkhorst}}\ and\ \bibinfo {author} {\bibfnamefont {J.~D.}\ \bibnamefont
  {Pack}},\ }\href {\doibase 10.1103/PhysRevB.13.5188} {\bibfield  {journal}
  {\bibinfo  {journal} {Phys. Rev. B}\ }\textbf {\bibinfo {volume} {13}},\
  \bibinfo {pages} {5188} (\bibinfo {year} {1976})}\BibitemShut {NoStop}%
\bibitem [{\citenamefont {Togo}\ and\ \citenamefont
  {Tanaka}(2015)}]{TOGO20151}%
  \BibitemOpen
  \bibfield  {author} {\bibinfo {author} {\bibfnamefont {A.}~\bibnamefont
  {Togo}}\ and\ \bibinfo {author} {\bibfnamefont {I.}~\bibnamefont {Tanaka}},\
  }\href {\doibase https://doi.org/10.1016/j.scriptamat.2015.07.021} {\bibfield
   {journal} {\bibinfo  {journal} {Scr. Mater.}\ }\textbf {\bibinfo {volume}
  {108}},\ \bibinfo {pages} {1 } (\bibinfo {year} {2015})}\BibitemShut
  {NoStop}%
\bibitem [{\citenamefont {Stokes}\ \emph {et~al.}()\citenamefont {Stokes},
  \citenamefont {Hatch},\ and\ \citenamefont {Campbell}}]{IsoTrop}%
  \BibitemOpen
  \bibfield  {author} {\bibinfo {author} {\bibfnamefont {H.~T.}\ \bibnamefont
  {Stokes}}, \bibinfo {author} {\bibfnamefont {D.~M.}\ \bibnamefont {Hatch}}, \
  and\ \bibinfo {author} {\bibfnamefont {B.~J.}\ \bibnamefont {Campbell}},\
  }\href@noop {} {\bibinfo  {journal} {ISOTROPY Software Suite
  [http://iso.byu.edu]}\ }\BibitemShut {NoStop}%
\bibitem [{\citenamefont {Cheng}\ \emph {et~al.}(2019)\citenamefont {Cheng},
  \citenamefont {Daemen}, \citenamefont {Kolesnikov},\ and\ \citenamefont
  {Ramirez-Cuesta}}]{doi:10.1021/acs.jctc.8b01250}%
  \BibitemOpen
\bibfield  {journal} {  }\bibfield  {author} {\bibinfo {author} {\bibfnamefont
  {Y.~Q.}\ \bibnamefont {Cheng}}, \bibinfo {author} {\bibfnamefont {L.~L.}\
  \bibnamefont {Daemen}}, \bibinfo {author} {\bibfnamefont {A.~I.}\
  \bibnamefont {Kolesnikov}}, \ and\ \bibinfo {author} {\bibfnamefont {A.~J.}\
  \bibnamefont {Ramirez-Cuesta}},\ }\href {\doibase 10.1021/acs.jctc.8b01250}
  {\bibfield  {journal} {\bibinfo  {journal} {J. Chem. Theory Comput.}\
  }\textbf {\bibinfo {volume} {15}},\ \bibinfo {pages} {1974} (\bibinfo {year}
  {2019})}\BibitemShut {NoStop}%
\bibitem [{\citenamefont {Parshall}\ \emph {et~al.}(2014)\citenamefont
  {Parshall}, \citenamefont {Heid}, \citenamefont {Niedziela}, \citenamefont
  {Wolf}, \citenamefont {Stone}, \citenamefont {Abernathy},\ and\ \citenamefont
  {Reznik}}]{PhysRevB.89.064310}%
  \BibitemOpen
  \bibfield  {author} {\bibinfo {author} {\bibfnamefont {D.}~\bibnamefont
  {Parshall}}, \bibinfo {author} {\bibfnamefont {R.}~\bibnamefont {Heid}},
  \bibinfo {author} {\bibfnamefont {J.~L.}\ \bibnamefont {Niedziela}}, \bibinfo
  {author} {\bibfnamefont {T.}~\bibnamefont {Wolf}}, \bibinfo {author}
  {\bibfnamefont {M.~B.}\ \bibnamefont {Stone}}, \bibinfo {author}
  {\bibfnamefont {D.~L.}\ \bibnamefont {Abernathy}}, \ and\ \bibinfo {author}
  {\bibfnamefont {D.}~\bibnamefont {Reznik}},\ }\href {\doibase
  10.1103/PhysRevB.89.064310} {\bibfield  {journal} {\bibinfo  {journal} {Phys.
  Rev. B}\ }\textbf {\bibinfo {volume} {89}},\ \bibinfo {pages} {064310}
  (\bibinfo {year} {2014})}\BibitemShut {NoStop}%
\bibitem [{\citenamefont {Ahmadova}\ \emph {et~al.}(2020)\citenamefont
  {Ahmadova}, \citenamefont {Sterling}, \citenamefont {Sokolik}, \citenamefont
  {Abernathy}, \citenamefont {Greven},\ and\ \citenamefont
  {Reznik}}]{Ahmadova2020}%
  \BibitemOpen
  \bibfield  {author} {\bibinfo {author} {\bibfnamefont {I.}~\bibnamefont
  {Ahmadova}}, \bibinfo {author} {\bibfnamefont {T.~C.}\ \bibnamefont
  {Sterling}}, \bibinfo {author} {\bibfnamefont {A.~C.}\ \bibnamefont
  {Sokolik}}, \bibinfo {author} {\bibfnamefont {D.~L.}\ \bibnamefont
  {Abernathy}}, \bibinfo {author} {\bibfnamefont {M.}~\bibnamefont {Greven}}, \
  and\ \bibinfo {author} {\bibfnamefont {D.}~\bibnamefont {Reznik}},\ }\href
  {\doibase 10.1103/PhysRevB.101.184508} {\bibfield  {journal} {\bibinfo
  {journal} {Phys. Rev. B}\ }\textbf {\bibinfo {volume} {101}},\ \bibinfo
  {pages} {184508} (\bibinfo {year} {2020})}\BibitemShut {NoStop}%
\bibitem [{\citenamefont {Kim}\ \emph {et~al.}(2018)\citenamefont {Kim},
  \citenamefont {Leiner}, \citenamefont {Park}, \citenamefont {Oh},
  \citenamefont {Sim}, \citenamefont {Iida}, \citenamefont {Kamazawa},\ and\
  \citenamefont {Park}}]{PhysRevB.97.201113}%
  \BibitemOpen
  \bibfield  {author} {\bibinfo {author} {\bibfnamefont {T.}~\bibnamefont
  {Kim}}, \bibinfo {author} {\bibfnamefont {J.~C.}\ \bibnamefont {Leiner}},
  \bibinfo {author} {\bibfnamefont {K.}~\bibnamefont {Park}}, \bibinfo {author}
  {\bibfnamefont {J.}~\bibnamefont {Oh}}, \bibinfo {author} {\bibfnamefont
  {H.}~\bibnamefont {Sim}}, \bibinfo {author} {\bibfnamefont {K.}~\bibnamefont
  {Iida}}, \bibinfo {author} {\bibfnamefont {K.}~\bibnamefont {Kamazawa}}, \
  and\ \bibinfo {author} {\bibfnamefont {J.-G.}\ \bibnamefont {Park}},\ }\href
  {\doibase 10.1103/PhysRevB.97.201113} {\bibfield  {journal} {\bibinfo
  {journal} {Phys. Rev. B}\ }\textbf {\bibinfo {volume} {97}},\ \bibinfo
  {pages} {201113} (\bibinfo {year} {2018})}\BibitemShut {NoStop}%
\bibitem [{\citenamefont {Park}\ \emph {et~al.}(2016)\citenamefont {Park},
  \citenamefont {Oh}, \citenamefont {Leiner}, \citenamefont {Jeong},
  \citenamefont {Rule}, \citenamefont {Le},\ and\ \citenamefont
  {Park}}]{PhysRevB.94.104421}%
  \BibitemOpen
  \bibfield  {author} {\bibinfo {author} {\bibfnamefont {K.}~\bibnamefont
  {Park}}, \bibinfo {author} {\bibfnamefont {J.}~\bibnamefont {Oh}}, \bibinfo
  {author} {\bibfnamefont {J.~C.}\ \bibnamefont {Leiner}}, \bibinfo {author}
  {\bibfnamefont {J.}~\bibnamefont {Jeong}}, \bibinfo {author} {\bibfnamefont
  {K.~C.}\ \bibnamefont {Rule}}, \bibinfo {author} {\bibfnamefont {M.~D.}\
  \bibnamefont {Le}}, \ and\ \bibinfo {author} {\bibfnamefont {J.-G.}\
  \bibnamefont {Park}},\ }\href {\doibase 10.1103/PhysRevB.94.104421}
  {\bibfield  {journal} {\bibinfo  {journal} {Phys. Rev. B}\ }\textbf {\bibinfo
  {volume} {94}},\ \bibinfo {pages} {104421} (\bibinfo {year}
  {2016})}\BibitemShut {NoStop}%
\bibitem [{\citenamefont {Sukhanov}\ \emph {et~al.}(2019)\citenamefont
  {Sukhanov}, \citenamefont {Pavlovskii}, \citenamefont {Bourges},
  \citenamefont {Walker}, \citenamefont {Manna}, \citenamefont {Felser},\ and\
  \citenamefont {Inosov}}]{PhysRevB.99.214445}%
  \BibitemOpen
  \bibfield  {author} {\bibinfo {author} {\bibfnamefont {A.~S.}\ \bibnamefont
  {Sukhanov}}, \bibinfo {author} {\bibfnamefont {M.~S.}\ \bibnamefont
  {Pavlovskii}}, \bibinfo {author} {\bibfnamefont {P.}~\bibnamefont {Bourges}},
  \bibinfo {author} {\bibfnamefont {H.~C.}\ \bibnamefont {Walker}}, \bibinfo
  {author} {\bibfnamefont {K.}~\bibnamefont {Manna}}, \bibinfo {author}
  {\bibfnamefont {C.}~\bibnamefont {Felser}}, \ and\ \bibinfo {author}
  {\bibfnamefont {D.~S.}\ \bibnamefont {Inosov}},\ }\href {\doibase
  10.1103/PhysRevB.99.214445} {\bibfield  {journal} {\bibinfo  {journal} {Phys.
  Rev. B}\ }\textbf {\bibinfo {volume} {99}},\ \bibinfo {pages} {214445}
  (\bibinfo {year} {2019})}\BibitemShut {NoStop}%
\bibitem [{\citenamefont {Sen}\ \emph {et~al.}(2019)\citenamefont {Sen},
  \citenamefont {Yao}, \citenamefont {Heid}, \citenamefont {Omoumi},
  \citenamefont {Hardy}, \citenamefont {Willa}, \citenamefont {Merz},
  \citenamefont {Haghighirad},\ and\ \citenamefont
  {Le~Tacon}}]{PhysRevB.100.104301}%
  \BibitemOpen
  \bibfield  {author} {\bibinfo {author} {\bibfnamefont {K.}~\bibnamefont
  {Sen}}, \bibinfo {author} {\bibfnamefont {Y.}~\bibnamefont {Yao}}, \bibinfo
  {author} {\bibfnamefont {R.}~\bibnamefont {Heid}}, \bibinfo {author}
  {\bibfnamefont {A.}~\bibnamefont {Omoumi}}, \bibinfo {author} {\bibfnamefont
  {F.}~\bibnamefont {Hardy}}, \bibinfo {author} {\bibfnamefont
  {K.}~\bibnamefont {Willa}}, \bibinfo {author} {\bibfnamefont
  {M.}~\bibnamefont {Merz}}, \bibinfo {author} {\bibfnamefont {A.~A.}\
  \bibnamefont {Haghighirad}}, \ and\ \bibinfo {author} {\bibfnamefont
  {M.}~\bibnamefont {Le~Tacon}},\ }\href {\doibase 10.1103/PhysRevB.100.104301}
  {\bibfield  {journal} {\bibinfo  {journal} {Phys. Rev. B}\ }\textbf {\bibinfo
  {volume} {100}},\ \bibinfo {pages} {104301} (\bibinfo {year}
  {2019})}\BibitemShut {NoStop}%
\bibitem [{\citenamefont {Yu}\ \emph {et~al.}(2015)\citenamefont {Yu},
  \citenamefont {Wu}, \citenamefont {Hu}, \citenamefont {Zhao}, \citenamefont
  {Li}, \citenamefont {Yang}, \citenamefont {Cheng}, \citenamefont {Luo},
  \citenamefont {Wang},\ and\ \citenamefont {Mao}}]{Yu14766}%
  \BibitemOpen
  \bibfield  {author} {\bibinfo {author} {\bibfnamefont {Z.}~\bibnamefont
  {Yu}}, \bibinfo {author} {\bibfnamefont {W.}~\bibnamefont {Wu}}, \bibinfo
  {author} {\bibfnamefont {Q.}~\bibnamefont {Hu}}, \bibinfo {author}
  {\bibfnamefont {J.}~\bibnamefont {Zhao}}, \bibinfo {author} {\bibfnamefont
  {C.}~\bibnamefont {Li}}, \bibinfo {author} {\bibfnamefont {K.}~\bibnamefont
  {Yang}}, \bibinfo {author} {\bibfnamefont {J.}~\bibnamefont {Cheng}},
  \bibinfo {author} {\bibfnamefont {J.}~\bibnamefont {Luo}}, \bibinfo {author}
  {\bibfnamefont {L.}~\bibnamefont {Wang}}, \ and\ \bibinfo {author}
  {\bibfnamefont {H.-k.}\ \bibnamefont {Mao}},\ }\href {\doibase
  10.1073/pnas.1520570112} {\bibfield  {journal} {\bibinfo  {journal} {Proc.
  Natl. Acad. Sci. U.S.A.}\ }\textbf {\bibinfo {volume} {112}},\ \bibinfo
  {pages} {14766} (\bibinfo {year} {2015})}\BibitemShut {NoStop}%
\end{thebibliography}%

\end{document}